 \documentclass[12pt,preprint]{aastex}

\pdfoutput=1
\usepackage{epstopdf}
\usepackage{placeins}
\usepackage{float}
\usepackage{subfigure}
\usepackage{amsmath}
\usepackage{textcomp}
\usepackage{gensymb}

\begin{document}

\title{RE-EXAMINING SUNSPOT TILT ANGLE TO INCLUDE ANTI-HALE STATISTICS}

\author{B. H. McClintock}
\affil{University of Southern Queensland, Toowoomba, 4350, Australia}
\email{u1049686@umail.usq.edu.au}

\author{A. A. Norton}
\affil{HEPL, Stanford University, Palo Alto, CA 94305, USA}
\email{aanorton@stanford.edu}

\and 

\author{J. Li}
\affil{Department of Earth, Planetary, and Space Sciences, University of California at Los Angeles, Los Angeles, CA 90095, USA}
\email{jli@igpp.ucla.edu}


\begin{abstract}
Sunspot groups and bipolar magnetic regions (BMRs) serve as an observational diagnostic of the solar cycle. We use Debrecen Photohelographic Data (DPD) from 1974-2014 that determined sunspot tilt angles from daily white light observations, and data provided by Li \& Ulrich (2012) that determined sunspot magnetic tilt angle using Mount Wilson magnetograms from 1974-2012.  The magnetograms allowed for BMR tilt angles that were anti-Hale in configuration, so tilt values ranged from 0 to $360$\degr\ rather than the more common $\pm90^{\circ}$. We explore the visual representation of magnetic tilt angles on a traditional butterfly diagram by plotting the mean area-weighted latitude of umbral activity in each bipolar sunspot group, including tilt information. The large scatter of tilt angles over the course of a single cycle and hemisphere prevents Joy's law from being visually identified in the tilt-butterfly diagram without further binning. The average latitude of anti-Hale regions does not differ from the average latitude of all regions in both hemispheres. The distribution of anti-Hale sunspot tilt angles are broadly distributed between 0 and $360$\degree\ with a weak preference for east-west alignment 180\degree\ from their expected Joy's law angle. The anti-Hale sunspots display a log-normal size distribution similar to that of all sunspots, indicating no preferred size for anti-Hale sunspots. We report that 8.4\%$\pm$0.8\% of all bipolar sunspot regions are misclassified as Hale in traditional catalogues. This percentage is slightly higher for groups within 5\degr\ of the Equator due to the misalignment of the magnetic and heliographic equators.
\end{abstract}


\section{Introduction}

\begin{figure} [!ht]
\centerline{\includegraphics[width=0.7\textwidth,clip=]{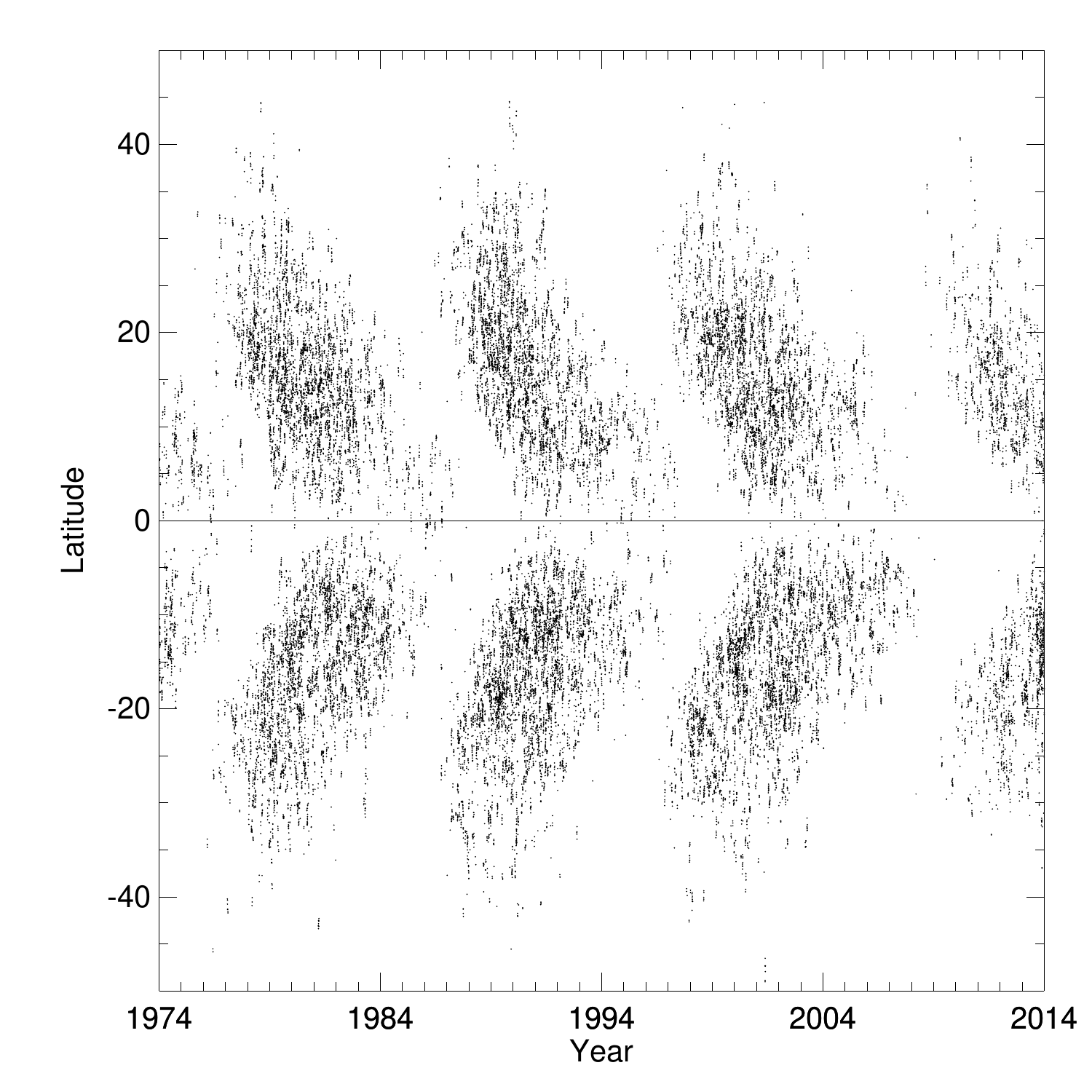}}
\caption{Latitude of bipolar sunspots as a function of time from daily DPD observations. Portions of Solar Cycles 20 and 24 are visible at the left and right edges of the diagram, respectively, while all of Cycles 21-23 are shown.}\label{bfly1}
 \end{figure}

The $\sim$ 11 yr pattern of sunspot activity begins at high latitudes (near $\pm 30^{\circ}$) to form a latitudinal band of activity.  The unsigned mean latitude of sunspot location decreases over time.  This is observed as a drift in the hemispheric latitudinal bands towards the equator later in the cycle. Late cycle sunspots finally appear near the equator while the next solar cycle sunspots begin emerging at high latitudes.   A diagram of sunspot latitude as a function of time over the course of a solar cycle resembles butterfly wings, as first noted by \citet{mau22}.  We show a version of the butterfly diagram in Figure \ref{bfly1} using the Debrecen Photohelographic Data (DPD) described in the next section.  \citet{car58} first noticed that the average latitude of sunspot emergence becomes increasing equatorward as the solar cycle progresses. \citet{mau22} said that the diagram ``seems to suggest three butterflies pinned down to a board with their wings extended.  Heads, bodies and legs have disappeared, but the outstretched wings remain.  Each pair of wings is distinct from the next; there is a clear V-shaped gap between each of the two specimens."

\begin{figure} [!ht]
\centerline{\includegraphics[width=1.0\textwidth,clip=]{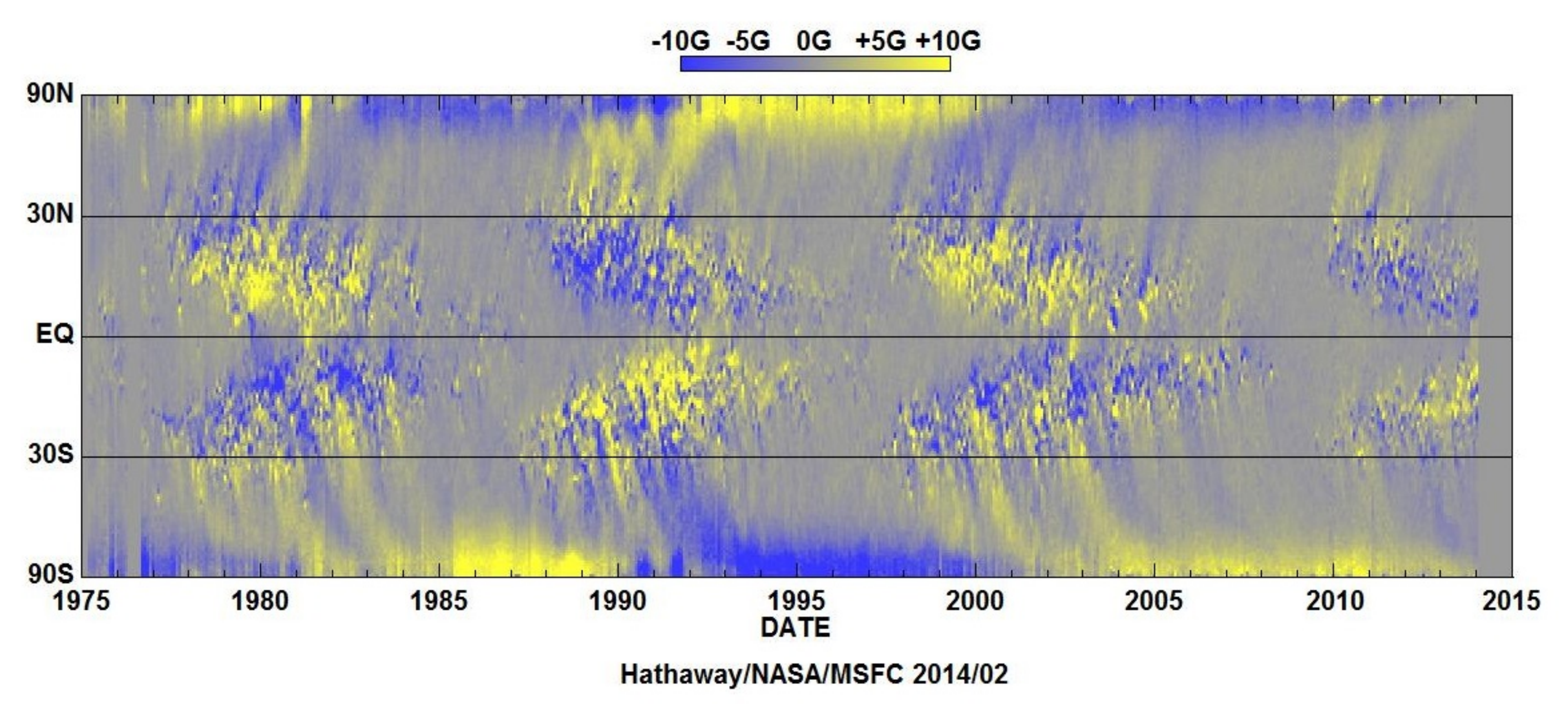}}
\caption{Diagram of the magnetic field of the Sun.  Yellow represents positive (outward) flux, blue represents negative (inward) flux. Hale's law can be observed. Courtesy of NASA/MSFC/David Hathaway.}\label{bflymag}
 \end{figure}

Observations of magnetic flux \citep{hal19} reveal sunspot groups have opposite polarities for leading and following spots with respect to solar rotation.  The majority of the time, the leading spot of a bipolar magnetic region (BMR) has the opposite polarity to leading spots in the other hemisphere.  With every solar cycle, the hemispheres alternate the dominant leading sunspot  polarity as seen in Figure \ref{bflymag}, courtesy of David Hathaway at NASA Marshall Space Flight Center.  Hale's law, as it is often called, denotes that if the northern hemisphere has a BMR configuration where the leading spot is positive and the following spot is negative, then the southern hemisphere would have the leading spot as negative and the following spot as positive.  BMRs that have the opposite orientation from the expected polarities are considered anti-Hale.

%

Leading spots, on average, are closer to the equator than following spots.  The difference in the latitudinal location of the leading and following spots is related as a tilt in the angle between the equator and a line drawn between bipolar sunspot groups.  Joy's law describes how BMRs on the solar surface are tilted with respect to the east-west equator of the Sun, with average tilt angle increasing as a function of increasing latitude.  Sunspot tilt angle may inform us about the process by which bipolar magnetic activity originates and rises to the surface.  The large scatter in tilt angles makes it difficult to recover Joy's law for a single solar cycle or individual hemisphere \citep{mcc13}. There is some indication that the slope of Joy's law is anti-correlated with the strength of a solar cycle \citep{das10}, although this is still under debate.

%

Historical data used for the determination of BMR tilt angle have not included magnetic polarity information to correctly identify anti-Hale regions.  Anti-Joy regions (those with the follower spot closer to the equator than the leading spot) are not necessarily anti-Hale \citep{tla13}. The existence of significant numbers of anti-Hale spots has been argued as evidence that our current understanding of sunspot formation is incomplete, if not flawed.  \citet{ste12} speculate that anti-Hale spots were caused by the existence of toroidal flux bands in opposite orientation at the same latitudinal in the interior.  

Tilt angle has been historically determined using white light observations from which magnetic polarity and anti-Hale information cannot be extracted. Observational studies of BMRs include polarity information, but many of the BMR studies include regions that are not sunspots since they do not have an umbra and penumbra seen in continuum intensity.  The BMRs in these studies include smaller active regions without sufficient flux to form sunspots or former sunspot regions that have broken apart.  One example is the work of \citet{wan89} who observed 2706 BMRs with fluxes $\ge 3 \times 10^{20}$ Mx from daily magnetograms obtained at the National Solar Observatory (NSO) during Solar Cycle 21 between 1976-1986.  Using the expected polarity for that cycle, they classified 113 BMRs as anti-Hale or approximately 4\%.  \citet{ste12} using MDI/SOHO magnetograms confirmed 4\% of mid-size to large BMRs between 1995-2011 were anti-Hale; however, smaller regions unlikely to form sunspots exceeded 25\%.  

Others have reported similar low percentages of anti-Hale regions; 3.1\% \citep{ric48} and $<$5\% \citep{smi67}.  However, these studies included in their total number of regions those that were not sunspots, regions that were unipolar, and poorly observed regions. (Note: tilt angles were not determined for the unipolar regions, but they were counted as part of the total number of regions to determine a percentage.)  \citet{khl09} noted that their determination of 4.9\% anti-Hale of bipolar sunspot regions might be low due to regions not properly recorded as anti-Hale in the data.  \citet{har92} used Mount Wilson sunspot polarity drawings and NSO full-disk magnetograms to propose that anti-Hale regions at high latitudes may be an indication of the next solar cycle starting earlier than previously established.  This is an obvious problem if a single date is used to separate one cycle from another.  We avoid this problem by separating the distinct cycles as a function of time and latitude so that high-latitude new cycle spots can be correctly identified and not mistaken for anti-Hale regions.  

\citet{mau04} noticed that the Southern hemisphere was producing spots on and across the Equator for Solar Cycle 12. He states ``Though the diagram shows clearly that there is but a single spot-zone in either hemisphere in each of these two cycles, a zone which moves in general accordance with Sp\"orer's curves, it reveals a striking and unexpected fact -- namely, that the southern current not only reaches the equator, but crosses it. The limit which bounds spot-distribution in the southern hemisphere on the equatorial side can be traced not only as far as the equator, but beyond it."  Sp\"orer's curves represent the progressive migration of sunspot mean latitude toward the Equator over the course of a solar cycle \citep{spo90}. Throughout this paper, we make a distinction between the magnetic and heliographic equators.  \citet{zol09} define the magnetic equator as the difference in latitude of sunspot production between the hemispheres.  We define magnetic equator in a similar fashion using only sunspot group latitudes. We explore whether a percentage of anti-Hale regions near the Equator are due to Northern polarity spots appearing below the heliographic equator or vice versa. 

\section{Data}

The Debrecen Photohelographic Data (DPD),\footnote{$http://fenyi.solarobs.unideb.hu/DPD/index.html$} spanning dates from 1974 January 2 to the present, consist of daily white light images taken primarily at the Heliophysical Observatory of the Hungarian Academy of Sciences \citep{gyo11}.  Our collection from 1974 January 2 to 2014 February 27 is based on 38,852 daily measurements of tilt angle and latitude (Figure \ref{bfly1}). We average latitude and tilt over the life of each region, resulting in 6968 sunspot regions. At the time of publication, data were still under preparation for 1980 to 1985 due to missing plates.  Note that the Hungarian Academy of Sciences houses a similar data set using  hourly MDI observations known as SOHO/MDI-Debrecen Data (SDD) with dates ranging from 1996 May 19 to 2010 December 31, although we are not utilizing the SDD data in this paper because they only contain one cycle and do not currently offer advantages over the DPD data other than an hourly cadence and coverage uninterrupted by poor atmospheric conditions.  The SDD data use the polarity of sunspots only ``to separate to the following or leading portion of the group independently from the geometrical position of spots" and do not indicate ``whether the polarities of the leading and following part follows the Hale' polarity law or not, the leading part is always that part which is in the leading position according to its longitude."\footnote{$ftp://fenyi.solarobs.unideb.hu/pub/SDD/additional/tilt\_angle/Readme.txt$}  The SDD data has the potential to report on Hale's polarity law and record a full $360^{\circ}$ range of tilt angles.  We hope this paper emphasizes the importance of anti-Hale statistics in order to encourage catalogues like SDD or STARA\footnote{$http://www.nso.edu/staff/fwatson/STARA/catalogue$} (another sunspot catalogue using MDI and HMI data) to include polarity information with full tilt angle ranges.

\citet{liu12} collected data from 1974 to 2012 using primarily Mount Wilson Observatory daily sunspot records and daily averaged magnetograms as well as MDI/SOHO magnetograms from 1996 to 2010.  Approximately 30,600 sunspot tilt angles were recorded with magnetic polarity information.  Only sunspot data were included, meaning there were no smaller magnetic regions that were not visible in white light images included in their sample.  Instead of plotting daily values, we average latitude and tilt over the life of each sunspot region, which provides a data set of 8377 bipolar sunspot regions.  See \citet{liu12} for all details and methodology, which include the assigning of ellipsoidal boundaries and centroids of polarity for active regions.

Magnetic information in historical data sets (Table \ref{tdata}) may have been used to identify that a certain spot group was bipolar, but not to establish a true magnetic tilt angle based on the dominant leading polarity for the hemisphere and solar cycle (i.e., tilt angles in historical data are limited to $\pm$90\degr\, not $0-360^{\circ}$).   Data of this type are flawed because anti-Hale regions are not recorded as such.  All data in Table 1 report tilt angles in the southern hemisphere as positive if the leading sunspot is closer to the equator than the following spot, regardless of the polarity of the leading sunspot.  It was only the combination of magnetic polarity information with sunspot data by \citet{liu12} that allows for the reporting and analysis of anti-Hale sunspot activity.

%
 \begin{table} [!ht]
 \begin{center}
 \caption{Data Containing Tilt Angle Measurements without Anti-Hale Information}\label{tdata}
 \begin{tabular}{lll}     
 \tableline\tableline
Data & Dates & Cadence\\
 \tableline
Mount Wilson (MW) & 1917-1985 & daily \\
Kodaikanal (KK) &1906-1987 & daily\\
Debrecen Photohelographic Data (DPD) &1974-2014& daily\\
SOHO/MDI-Debrecen Data (SDD) &1996-2010& hourly\\
SDO/HMI-Debrecen Sunspot Data (HMIDD) &2010-2013& hourly\\
 \tableline
 \end{tabular}
 \end{center}
\end{table}

\section{Adding Tilt Angle to Butterfly Diagrams}

\begin{figure}[!ht]
\centerline{\includegraphics[width=0.95\textwidth,clip=]{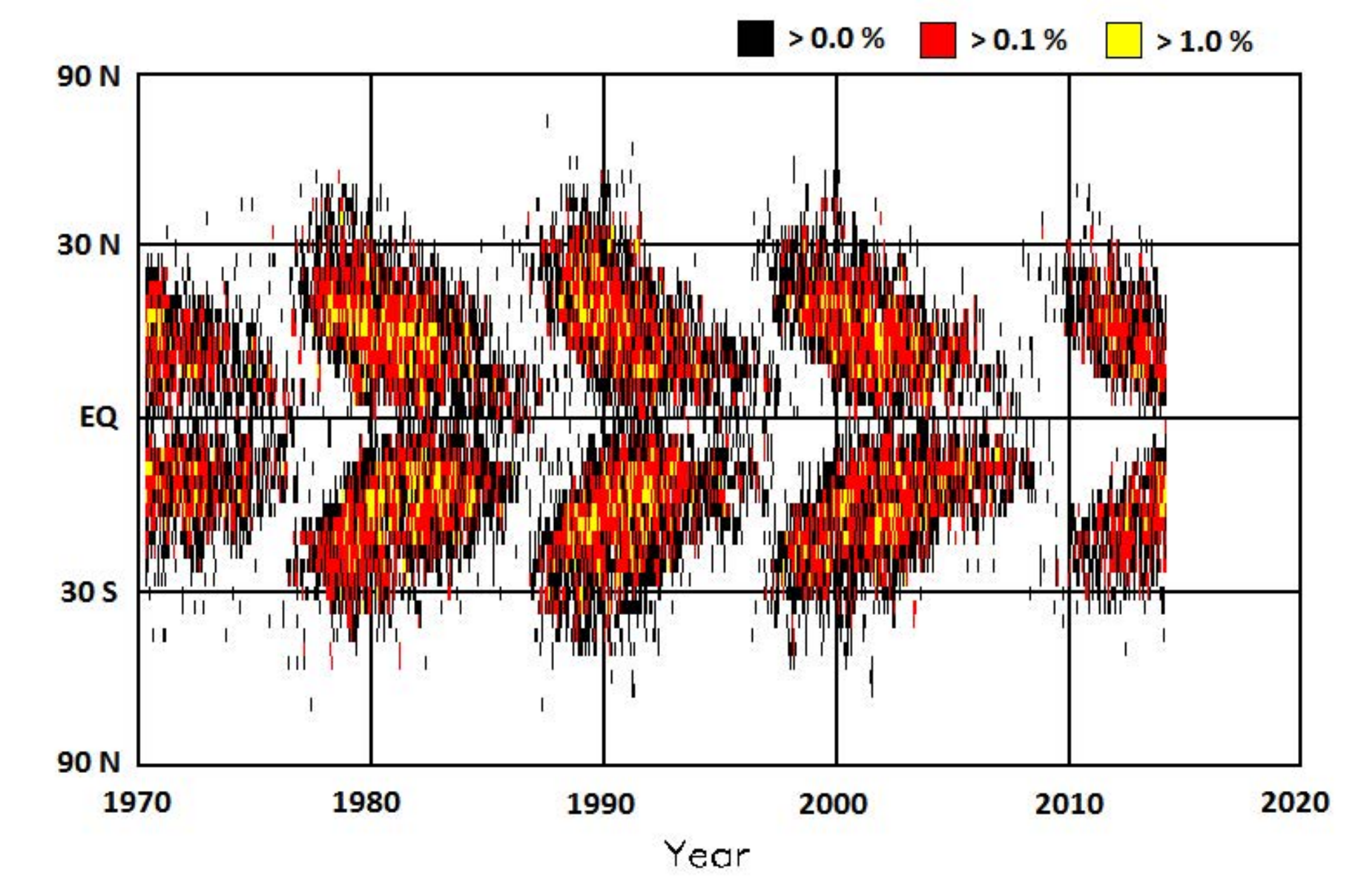}}
\caption{Daily mean sunspot area per each solar rotation plotted as a function of time and latitude.  Areas are binned in 50 equal-area latitude strips.  The relative area of the sunspot group is illustrated with black, red, and yellow as areas of increasing size.  Courtesy of NASA/MSFC/David Hathaway}\label{nasabfly2}
 \end{figure}

Variations in the way the butterfly diagram is plotted can illustrate characteristics of the solar cycle beyond the simple fact that sunspots move equatorward over time. For example, if sunspot area is included, as shown in Figure \ref{nasabfly2} and originally produced by \citet{hat03}, then details regarding the times of greatest sunspot area production are evident roughly in the center of the butterfly wings.  \citet{ter07,ter10} further studied the density of sunspot area as a function of time and latitude and found that in any hemisphere the activity is split into two or more distinct activity waves drifting equatorward. We explore whether the depiction of the bipolar sunspot region tilt angles as plotted in the classical butterfly diagram format can tell us anything more about the solar dynamo. 

%

\begin{figure} [!ht]
\centerline{\includegraphics[width=1.00\textwidth,clip=]{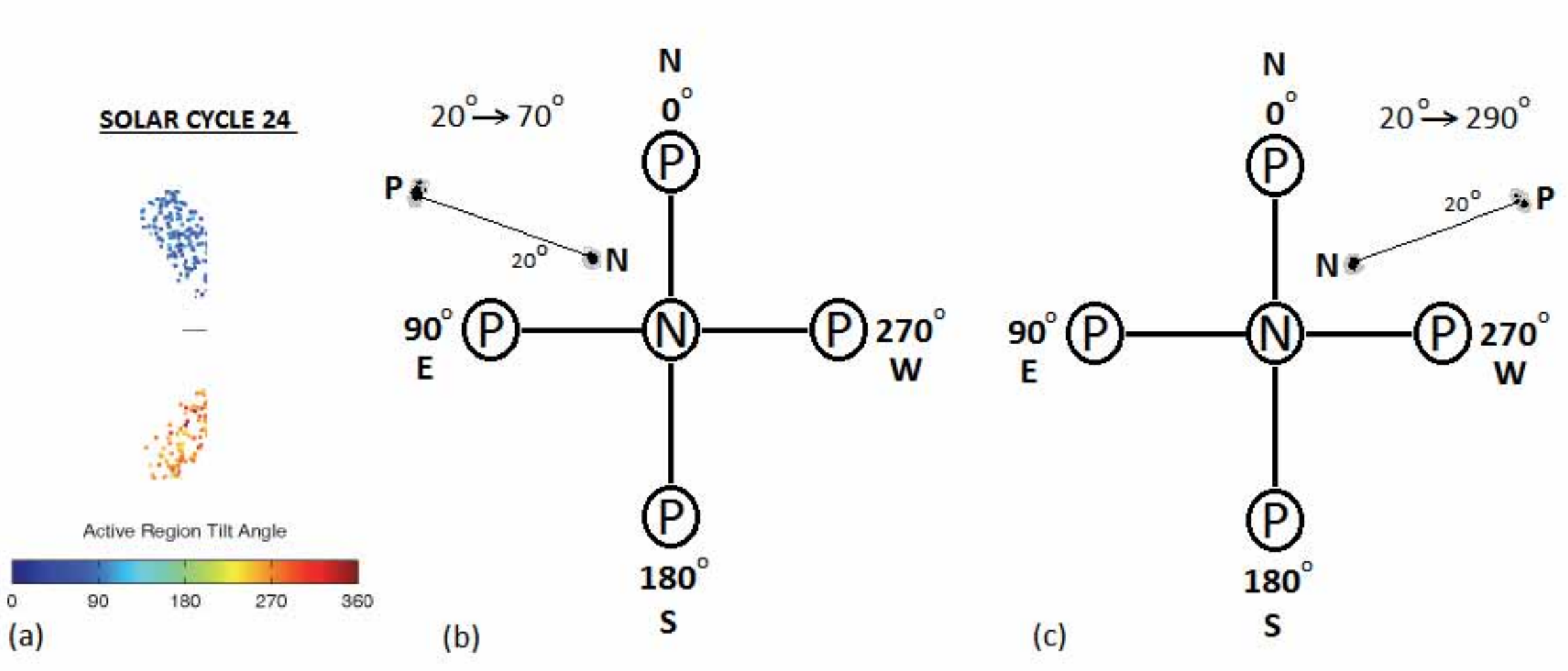}}
\caption{``P" is positive polarity, ``N" is negative. (a) An example of expected tilt angles for Solar Cycle 24.  We would expect tilt angles to average slightly less than 90$^{\circ}$ and slightly more than 270$^{\circ}$ for this cycle.  (b) The dominant leading polarity in the northern hemisphere for Solar Cycle 24 is negative.  A northern hemisphere tilt angle of 20$^{\circ}$ from the equator would be 70$^{\circ}$. (c) A southern hemisphere tilt angle of 20$^{\circ}$ from the equator would be 290$^{\circ}$. \citet{liu12} defined tilt angle from the west to produce a similar range of 360$^{\circ}$.}\label{sc24ex}
 \end{figure}

Using magnetic polarity information, \citet{liu12} defined tilt angle from the west to produce a range of -180$^{\circ}$ to 180$^{\circ}$. We define tilt angle in similar fashion, measuring tilt angle counterclockwise from the north to produce a full range of 360$^{\circ}$ as well (Figure \ref{sc24ex}), rather than the traditional tilt angle range of -90$^{\circ}$ to 90$^{\circ}$ used in other data sets, such as those listed in Table \ref{tdata}.  Our tilt angles have the same range as Li \& Ulrich, however by comparison we would define an angle of 90$^{\circ}$ by Li \& Ulrich as 0$^{\circ}$ in our orientation from the north.

Examples of expected tilt angles for Solar Cycle 24 are included in Figure \ref{sc24ex} for reference.  We define anti-Hale tilt angles in Solar Cycles 20, 22, and 24 for the northern (southern) hemisphere as between 0$^{\circ}$ and 180$^{\circ}$ (180$^{\circ}$ and 360$^{\circ}$). For Cycles 21 and 23, anti-Hale in the northern (southern) hemisphere is between 180$^{\circ}$ and 360$^{\circ}$ (0$^{\circ}$ and 180$^{\circ}$).  We would expect tilt angles in Cycle 24 for example to average slightly less than 90$^{\circ}$ in the Northern hemisphere and slightly more than 270$^{\circ}$ in the Southern hemisphere according to our definition of tilt angle.

%

\FloatBarrier

\begin{figure}[!ht]
\centerline{\includegraphics[width=0.7\textwidth,clip=]{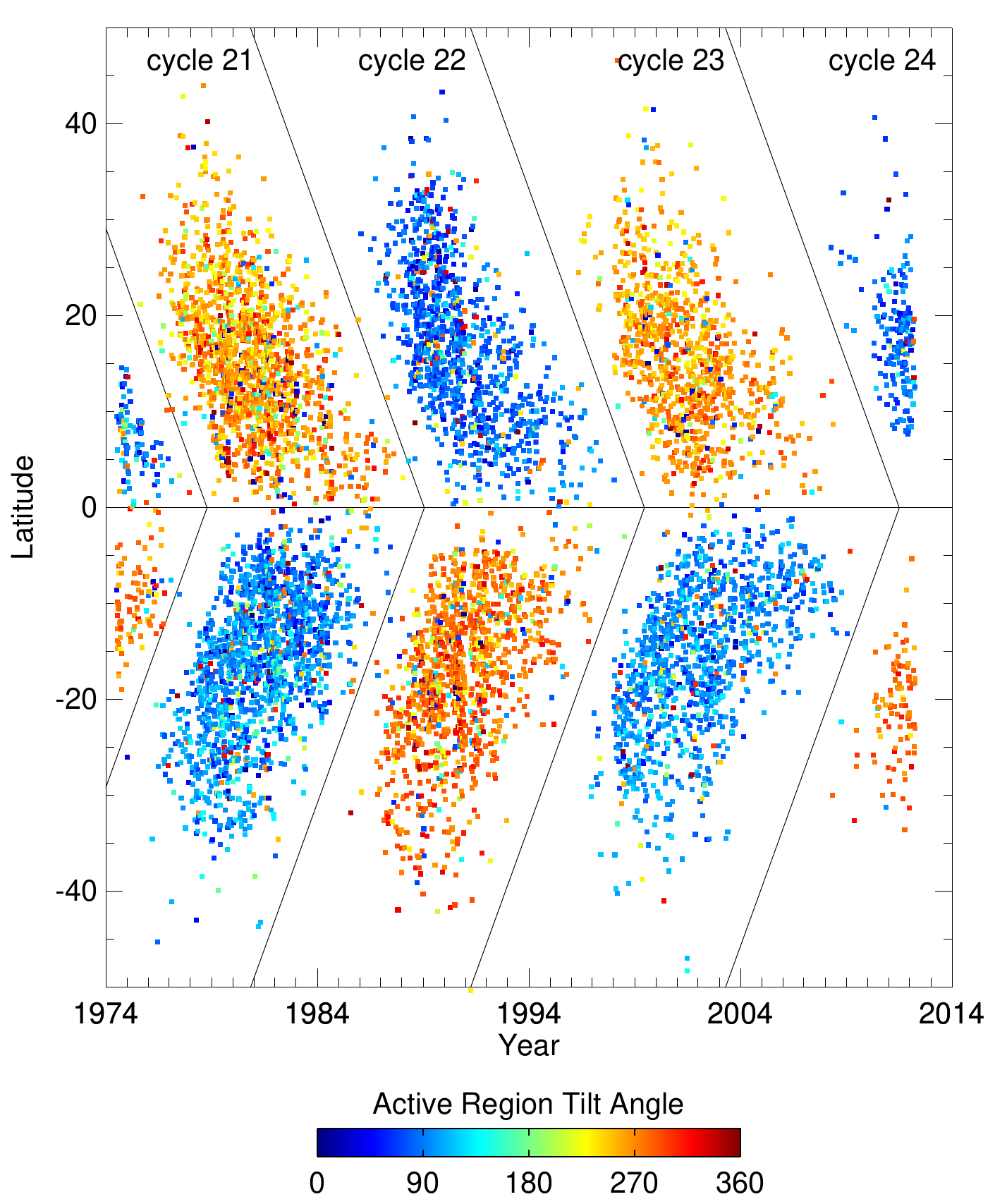}}
\caption{Tilt angles from \citet{liu12} data averaged by bipolar sunspot region and plotted as a function of time and latitude.  Tilt measured counterclockwise from the north. Slope $(\pm\frac{1}{60})$ and location of solar cycle boundaries chosen manually for best fit.  A portion of Solar Cycle 20 is visible at the left edge of the diagram.}\label{bflyli}
 \end{figure}

Solar Cycles 20 to 24 are plotted from Li \& Ulrich data to include tilt angle as measured counterclockwise from the north (Figure \ref{bflyli}).  The latitude of the sunspot group is an average of area-weighted latitude determinations of leading and following sunspot umbrae. Solar Cycles 20, 22, and 24 have a negative (positive) leading sunspot in the northern (southern) hemisphere.  Solar Cycles 21 and 23 are reversed to where the northern (southern) hemisphere has a positive (negative) leading sunspot.  

%

\FloatBarrier

The boundary between the hemispheres is defined by the heliographic equator and the boundaries between cycles for the southern/northern hemispheres are defined by $60 \times ylat = \pm (xdate - xint)$, where $ylat$ is latitude, $xdate$ is a function of the number of days from 1974 January 1 and $xint$ is the point of intersection at the equator.\footnote{$xdate=365.25(year-1974)+DOY$}  We visually determined the slope, $\frac{1}{60}$, as a value that could easily demarcate all of the cycles while $xint$ was visually determined separately for the best fit for each cycle as determined by the plot of all regions.  Values and dates are listed in Table \ref{tbound}. We explicitly denote the solar cycle boundaries in the paper because the determination of anti-Hale regions depends on it.

%
 \begin{table}[!ht]
\begin{center}
 \caption{Solar Cycle Boundary Information Used to Determine Anti-Hale Activity}\label{tbound}
 \begin{tabular}{clll}     
\tableline\tableline
Solar Cycle & Minimum & $xint$ & Date\\
 \tableline
20 & ... & 1 & 1974 Jan 1\\
20/21 & 1976 Jun & 1750 & 1978 Oct 16\\
21/22 & 1986 Sep & 5500 & 1989 Jan 21\\
22/23 & 1996 May & 9300 & 1999 Jun 18\\
23/24 & 2008 Jan & 13700 & 2011 Jul 5\\
\tableline
 \end{tabular}
\end{center}
 \end{table}

\begin{figure} [!ht]
\centerline{\includegraphics[width=0.70\textwidth,clip=]{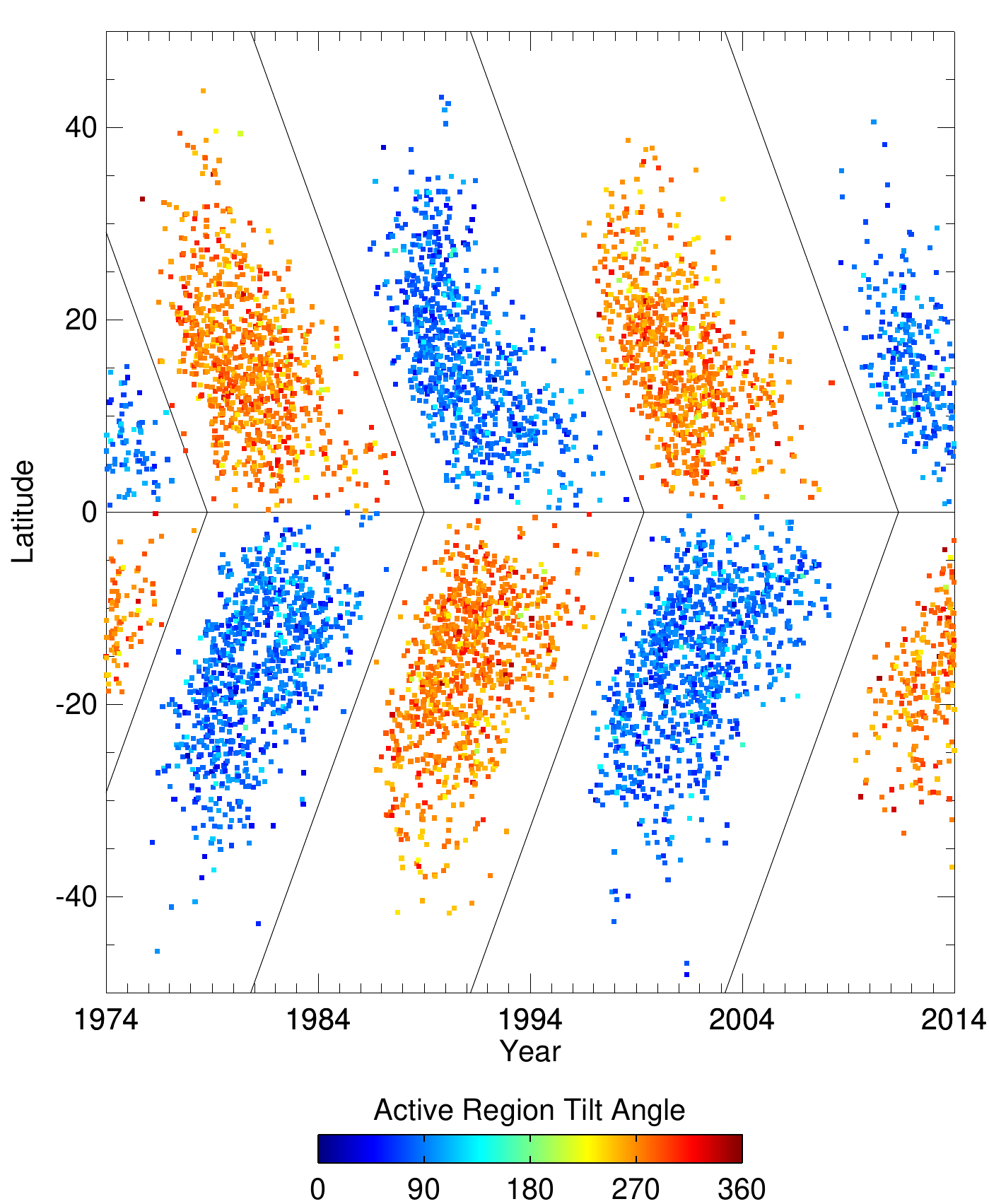}}
\caption{Tilt angles from DPD data averaged by bipolar sunspot region and plotted as a function of time and latitude.  Tilt measured counterclockwise from the north and assigned a range of either 0-180$^{\circ}$ or 180$^{\circ}$-360$^{\circ}$ based on the expected polarity orientation for each hemisphere in that solar cycle.  Portions of Solar Cycle 20 and 24 are visible at the left and right edges of the diagram, respectively, with all of Cycles 21-23 presented.}\label{bflydpd}
 \end{figure}

Solar Cycles 20 to 24 are plotted from DPD data to include tilt angle as measured counterclockwise from the north (Figure \ref{bflydpd}).  Latitude of the sunspot group is an average of area-weighted latitude determinations of leading and following sunspot umbrae. Sunspot groups near the equator are assigned a hemisphere as determined by mean latitude, regardless of leading spot polarity.  DPD data are limited to a tilt angle range of 180$^{\circ}$, so we used the expected polarity orientation for that solar cycle to assign a range of 0 to 180$^{\circ}$ to the appropriate hemisphere and 180$^{\circ}$ to 360$^{\circ}$ to the other hemisphere. Anti-Hale information is therefore not represented in either hemisphere for any solar cycle.

%

We use mean and median of tilt angles plotted over sunspot butterfly diagrams to confirm that tilt angle time dependence is a function of sunspot latitude \citep{liu12}.  We first binned all data points in 300 day intervals, then found mean and median tilt angles ($\gamma$)  of each bin using

 \begin{eqnarray} 
     mean(\gamma) = arctan \left( \frac{\sum sin\gamma}{\sum cos\gamma} \right) \,,      \label{mean} \\ 
     median(\gamma) = arctan \left( \frac{median(sin\gamma)}{median(cos\gamma)} \right) \,.  \label{median}    
  \end{eqnarray}

\begin{figure}  [!ht]
\centering
\subfigure[]{\includegraphics[width=0.65\textwidth,clip=]{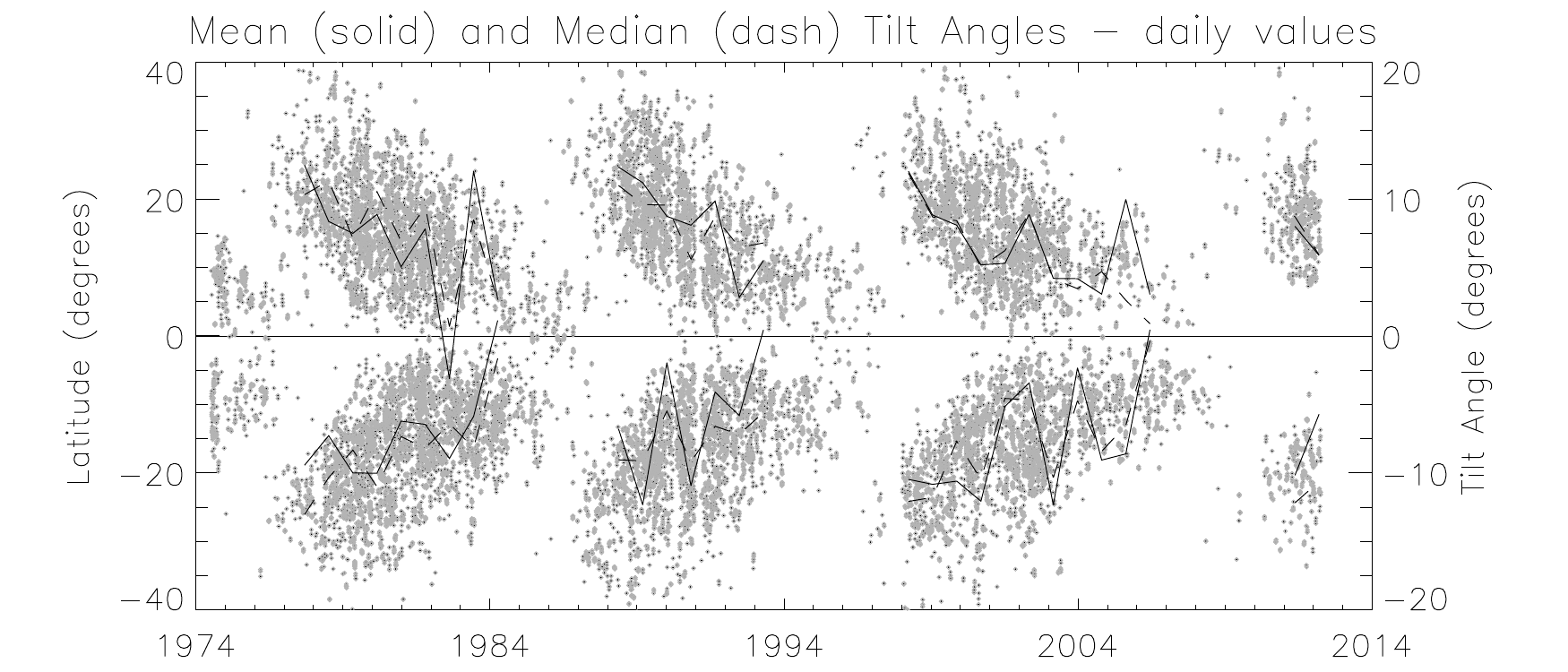}}
\subfigure[]{\includegraphics[width=0.65\textwidth,clip=]{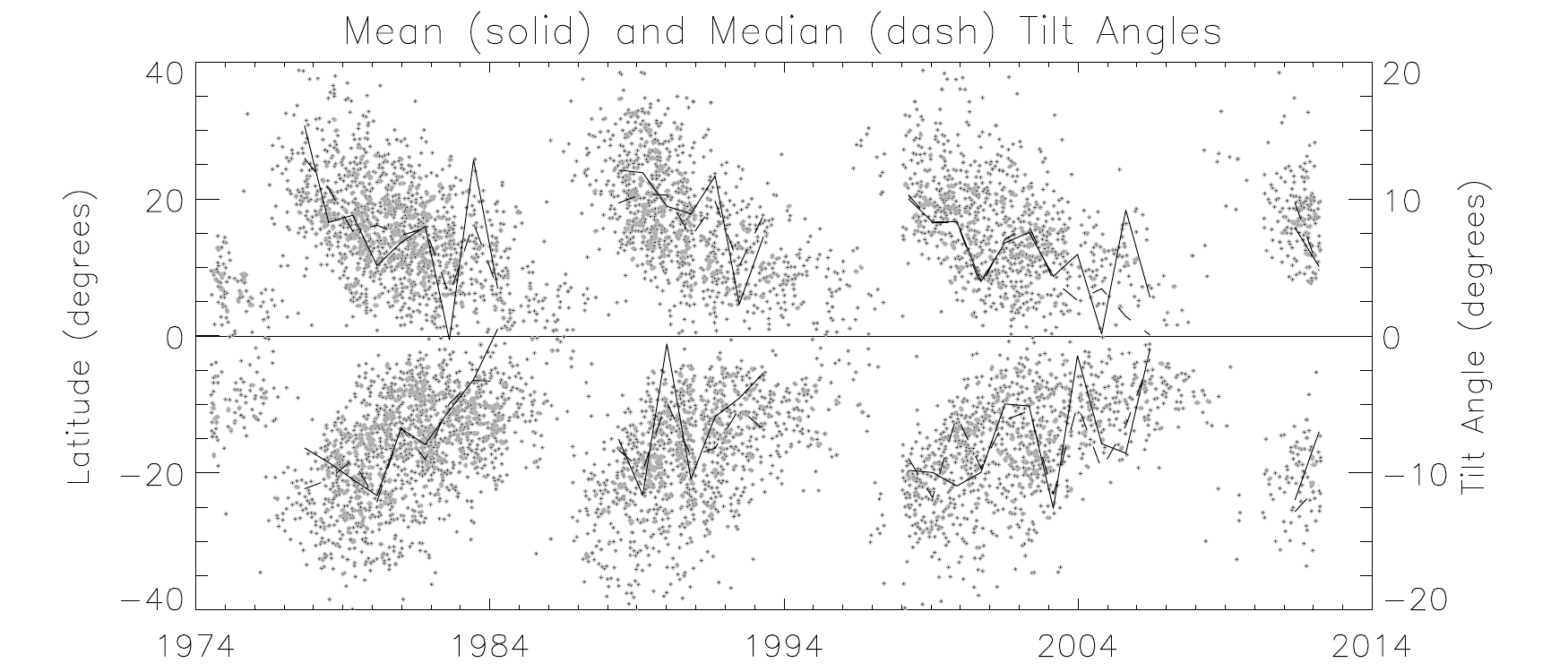}}
\subfigure[]{\includegraphics[width=0.65\textwidth,clip=]{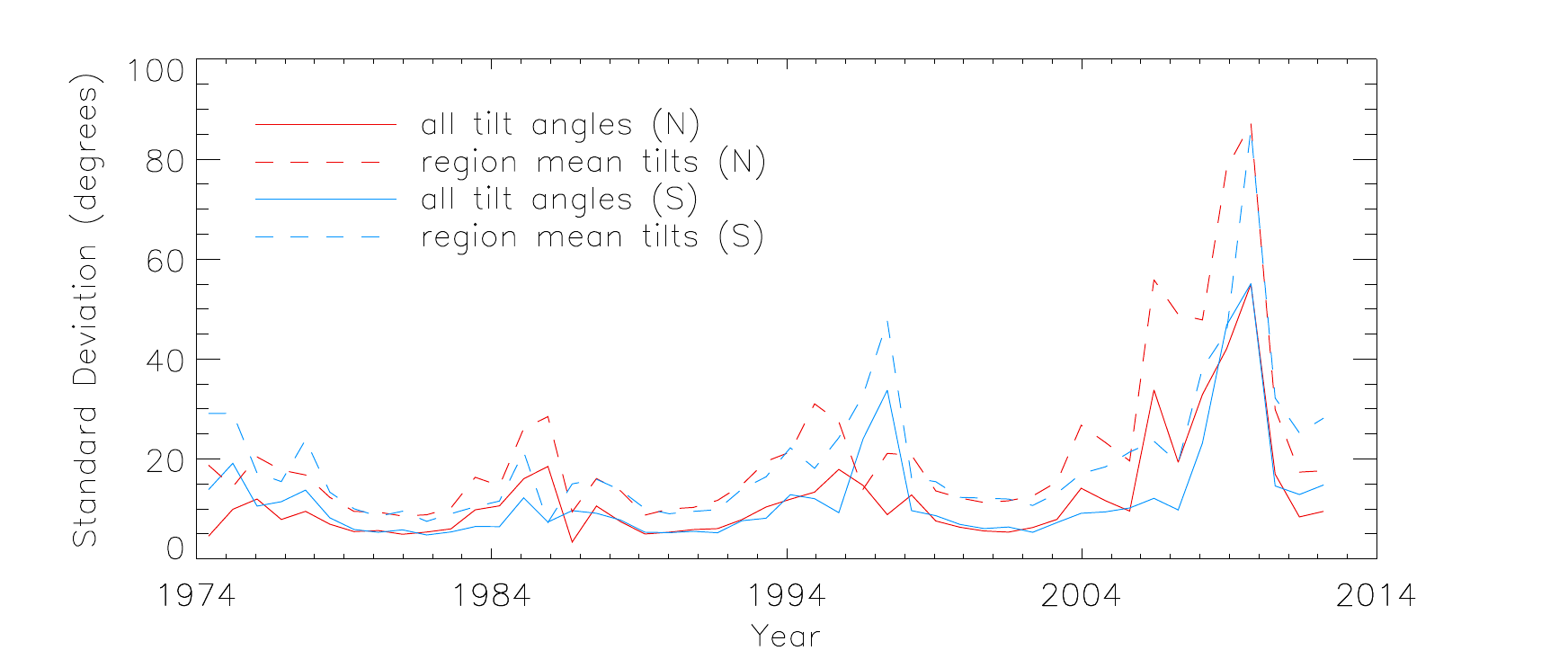}}
\caption{In panels (a) and (b), mean (solid) and median (dash) tilt angles binned in 300 day intervals are plotted over the sunspot butterfly diagram. Latitudes (tilt angles) are given in the left (right) vertical axis.  The horizontal straight line indicates the equator. In panel (a), sunspot tilt angles and latitudes are recorded for each day (daily) that the sunspot group is present on the disk. In panel (b), the average tilt and latitude for that group is recorded only once during its passage across the disk. In panel (c), standard deviation of all sunspot observations (solid) and active region means (dash) for the northern (red) and southern (blue) hemispheres are shown.} \label{mmli}
 \end{figure}

Mean (solid) and median (dashed) of each bin are plotted over the butterfly diagram of all sunspots in Figure \ref{mmli}(a). We also determined mean latitude and tilt angle for each active region, binned those values in 300-day intervals, and found mean and median of each bin (Figure \ref{mmli}(b)) using Equations \eqref{mean} and \eqref{median}. The sunspot latitude is given on the left vertical axis with a range of 0-40$^{\circ}$.  The tilt angle mean and median (lines) are shown on the right vertical axis with a range of 0-20$^{\circ}$. It should be noted that the tilt angles were not separated by cycle boundaries in Figure \ref{mmli}.  The 300-day bins near solar minimum will contain overlapping cycles for data shown in these figures.  Mean or median tilt angle is not useful near solar minimum and were therefore excluded from the plots.  Tilt angle decreases as sunspots migrate from high to low latitudes in each solar cycle, as expected from Joy's law. Note that the decrease of tilt angle over time is not smooth, but discontinuous, possibly an indication of the distinct dynamo waves as mentioned by \citet{ter07,ter10}. Standard deviation, $s(\gamma)$, as set by,

\begin{equation}
s(\gamma) = \sqrt{\frac{(\overline{sin\gamma})^{2}+(\overline{cos\gamma})^{2}}{n}}
\end{equation}

 of all sunspot observations (solid) and active region means (dash) for the Northern (red) and Southern (blue) hemispheres are shown in Figure \ref{mmli}(c). The peak of tilt angle scatter occurs as expected between solar cycles when polarities in both hemispheres are changing 180$^{\circ}$ in accordance with Hale's law. 

%

\section{Anti-Hale Regions}

\begin{figure}[!ht]
\centerline{\includegraphics[width=0.7\textwidth,clip=]{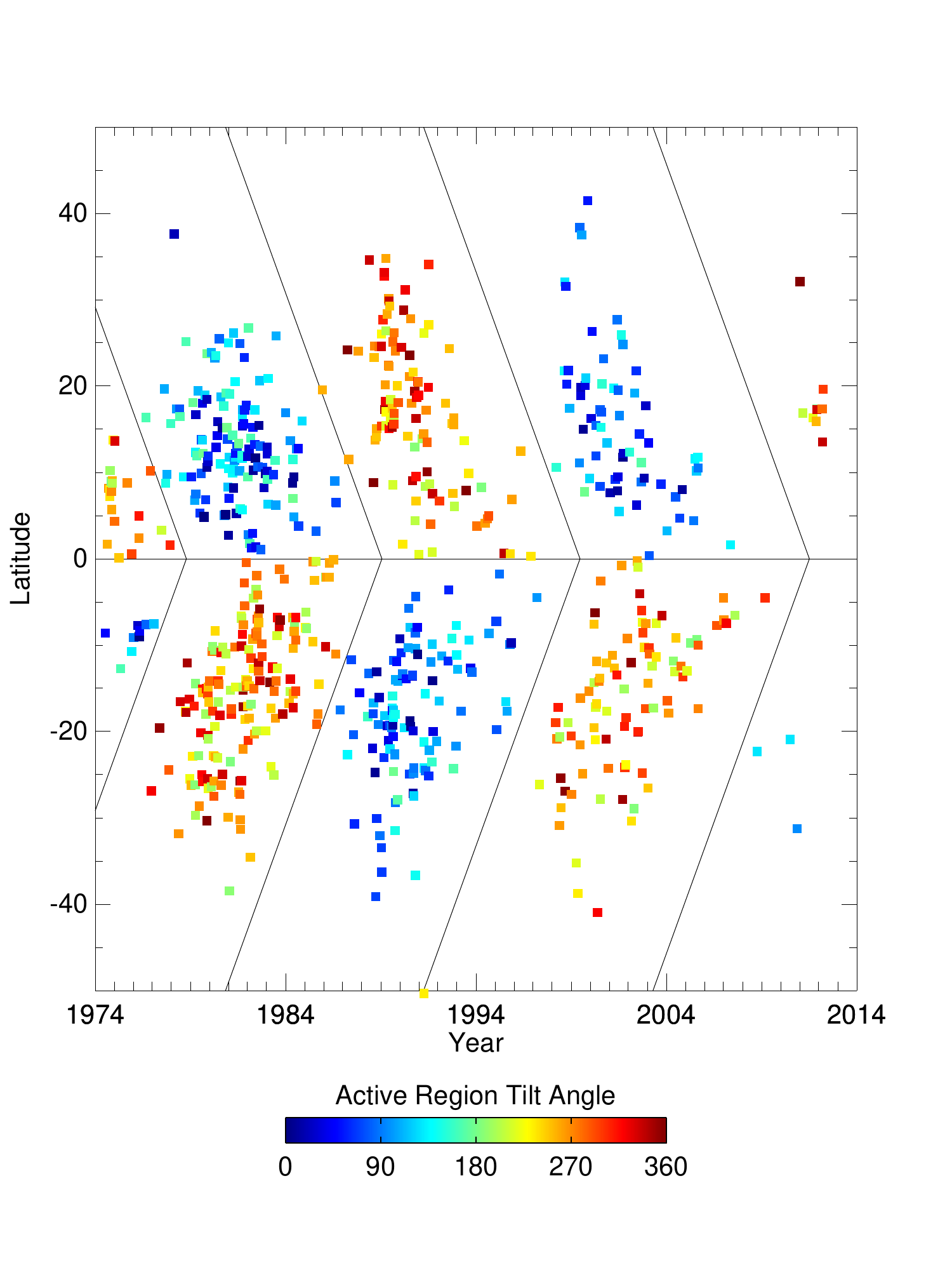}}
\caption{Anti-Hale tilt angles from \citet{liu12} data averaged by bipolar sunspot region and plotted as a function of time and latitude.  Tilt measured counterclockwise from the north.  Portions of Solar Cycle 20 and 24 are visible at the left and right edges of the diagram, respectively, with all of Cycles 21-23 presented. With fewer data points, larger pixel size assigned to make color variations more visible.}\label{bflyliah}
 \end{figure}

Li \& Ulrich data are used to plot anti-Hale bipolar sunspot regions from 1974 to 2012 as a function of time and latitude (Figure \ref{bflyliah}).  Tilt angles were averaged over the lifetime of each region. With fewer data points, a larger pixel size than previous figures is assigned to make color variations more visible.

%

The percentage of bipolar sunspot regions that are anti-Hale from 1974 to 2012 are shown in Table \ref{tsc}. See Table \ref{tbound} for solar cycle boundary definitions. The total sunspot area by hemisphere for Solar Cycles 20 to 23 from the archives of the Greenwich Royal Observatory\footnote{$http://solarscience.msfc.nasa.gov/greenwch.shtml$} is in 10$^4$ micro-hemispheres.  Sunspot area is used as a proxy for cycle strength \citep{sol93}.    Solar Cycles 20 and 24 only include partial bipolar sunspot data at the end and beginning of those cycles, respectively, which could explain the high percentage of anti-Hale in the northern hemisphere for Solar Cycle 20.  Of 8377 bipolar sunspot regions, 705 (8.4\%) were anti-Hale.  We found anti-Hale percentages of 9.0\%, 8.7\%, and 7.2\% for Solar Cycles 21, 22, and 23 respectively, with 14.8\% for Cycle 20 and 4.0\% for Cycle 24.

\FloatBarrier

%
 \begin{table}[!ht]
\begin{center}
\caption{Anti-Hale (AH) Information by Solar Cycle and Hemisphere with Total Area}\label{tsc}
\begin{tabular}{lcccccccc}     
\tableline\tableline
Cycle (yr) & Year of & \multicolumn{3}{c}{Northern Hemisphere} & \multicolumn{3}{c}{Southern Hemisphere}&Total\\
Length & Minimum & N & AH & $\Sigma$ Area\tablenotemark{a} & N & AH & $\Sigma$ Area\tablenotemark{a} & AH\\
 \tableline
20 (partial) & (1964.8) & (107) & (19.6\%) & 6.94 & (96) & (9.4\%) & 4.91 & 14.8\% \\
21 (10.3) & 1976.5 & 1547 & 8.4\% & 7.51 & 1710 & 9.5\% & 7.77 & 9.0\% \\
22 (10.0) & 1986.8 & 1150 & 9.0\% & 6.38 & 1287 & 8.5\% & 7.24 & 8.7\%\\
23 (12.2) & 1996.9 & 1004 & 6.6\% & 5.61 & 1178 & 7.8\% & 6.45 & 7.2\% \\
24 (partial) & 2008.1 & (193) & (4.7\%) & ... & (105) & (2.9\%) & ... & 4.0\% \\
\hline
Total & ... & 4001 & 8.2\% & ... & 4376 & 8.6\% & ... & 8.4\% \\
\tableline
 \end{tabular}
\tablenotetext{a}{total sunspot area in 10$^4$ micro-hemispheres.}
\end{center}
 \end{table}  

\begin{figure} [!ht]
\centering
\subfigure[]{\includegraphics[width=0.5\textwidth,clip=]{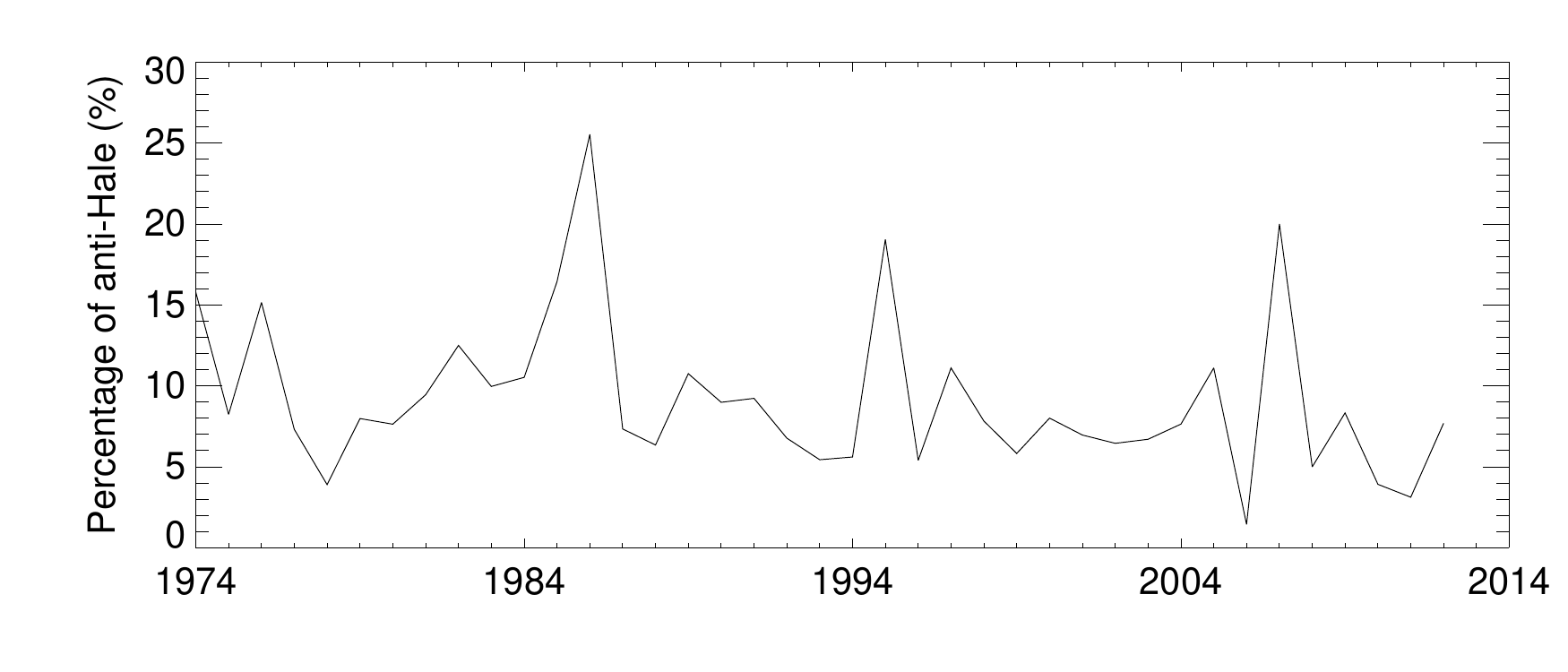}}
\subfigure[]{\includegraphics[width=0.5\textwidth,clip=]{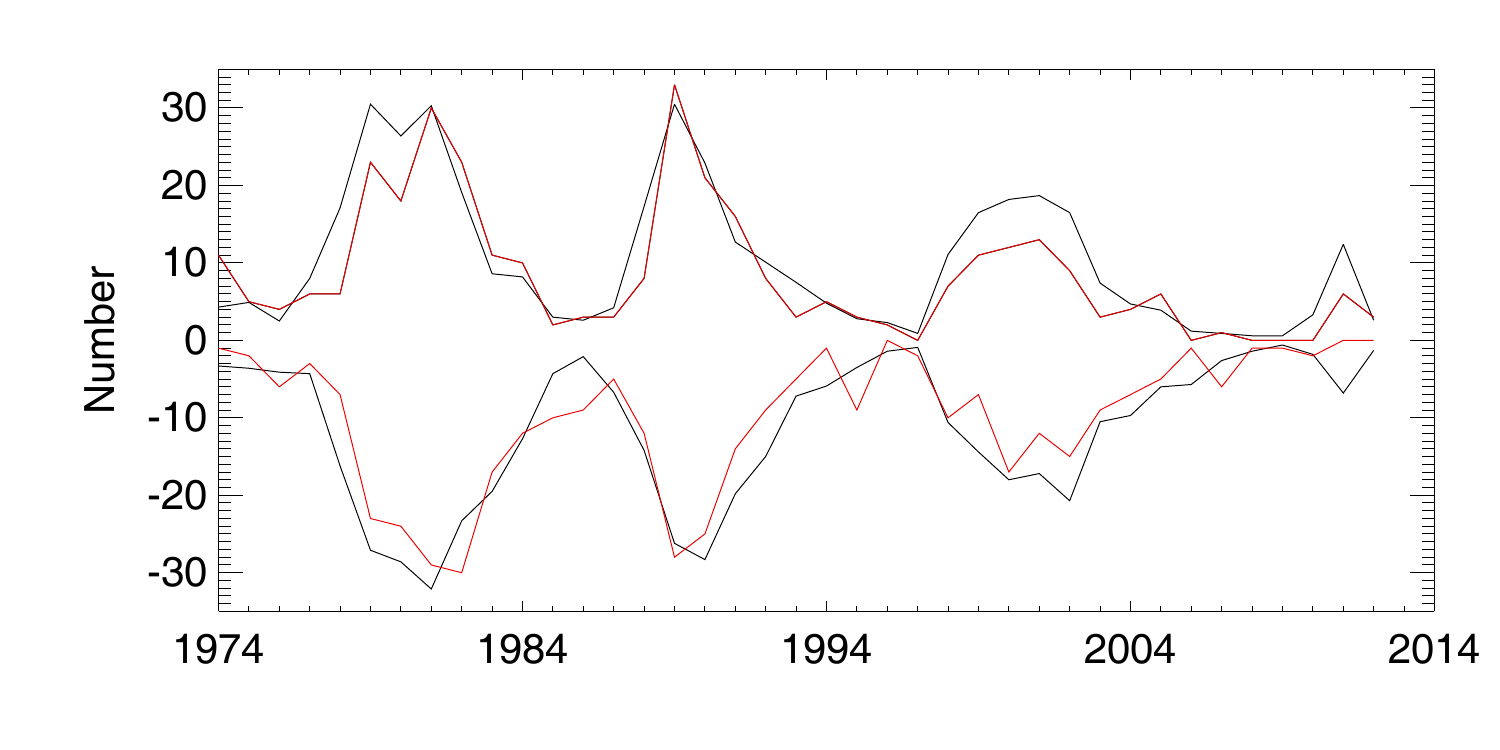}}
\subfigure[]{\includegraphics[width=0.5\textwidth,clip=]{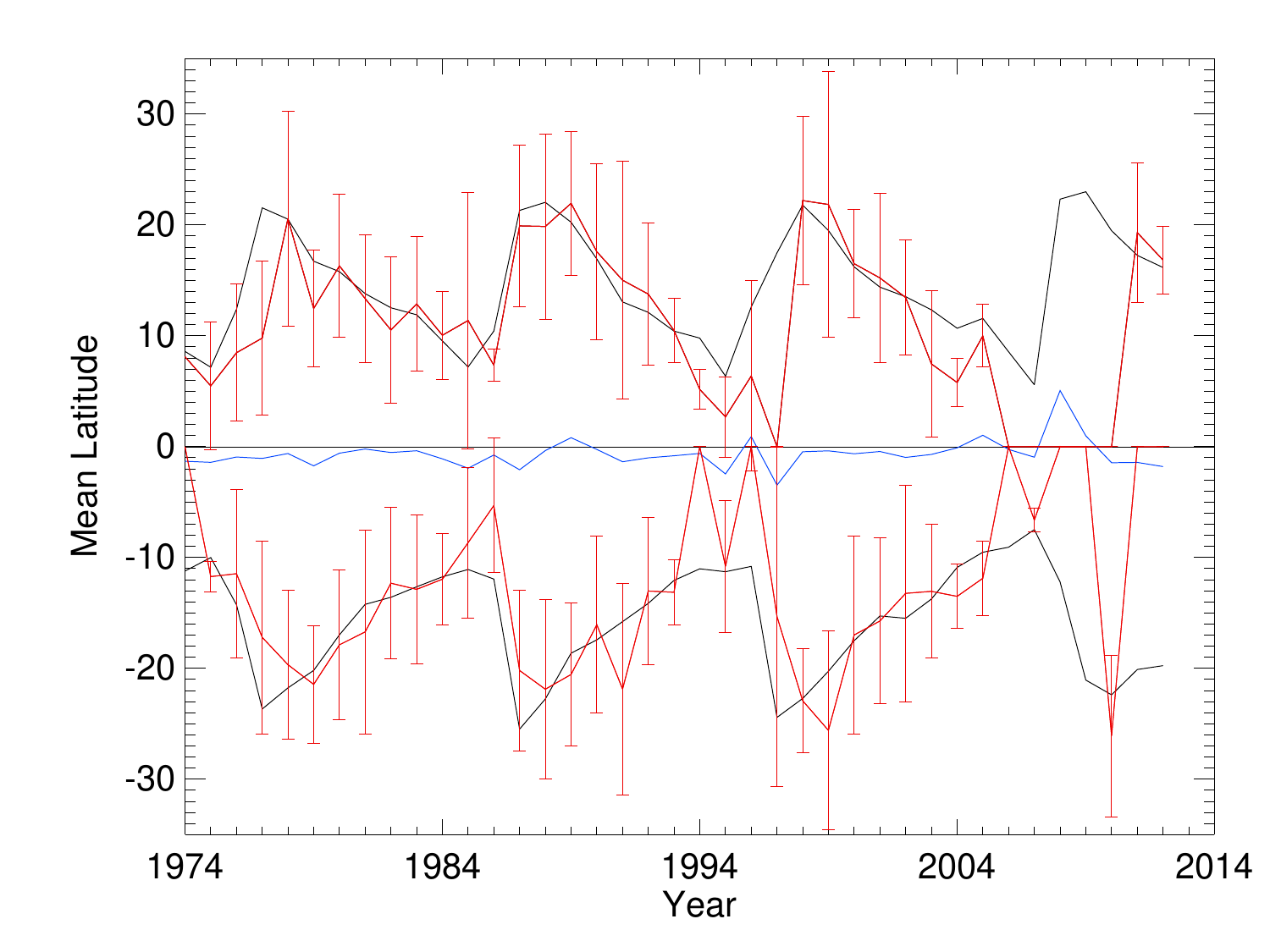}}
\caption{Yearly binning of \citet{liu12} data plotted over time.  (a) The percentage of bipolar sunspot regions that are anti-Hale. (b) The number of bipolar sunspot regions divided by 10 (black) and anti-Hale regions (red). (c) Mean latitude of bipolar sunspot regions (black) and anti-Hale (red). Standard deviation of anti-Hale bins plotted as error bars. The magnetic equator (blue) is defined as the difference in the latitudinal centroids of the sunspot locations in the hemispheres.  Note that the magnetic equator is deflected southward of the heliographic equator at all times except for the beginning of Solar Cycle 24.}\label{ahtot}
 \end{figure}

Figure \ref{ahtot}(a) shows that the percentage of anti-Hale regions binned yearly are relatively consistent over time except near the end of each cycle. This could be a result of activity occurring at low latitudes, thus interacting across the equator. The number of anti-Hale regions (red) closely tracks the number of bipolar sunspot regions divided by 10 (black) in Figure \ref{ahtot}(b).  Similar tracking occurs when mean latitude of bipolar sunspot regions (black) and anti-Hale (red) are plotted, with the standard deviation as error bars (Figure \ref{ahtot}(c)).  \citet{zol09} defined the magnetic equator as the difference in the latitudinal centroids of the sunspot locations in the hemispheres.  Our data are limited to sunspot groups, of which we take the yearly mean latitude in each hemisphere and average the two values to define the magnetic equator (blue). Note that the magnetic equator is deflected southward at all times except for the beginning of Solar Cycle 24.  \citet{zol09} also calculated the magnetic equator from Royal Greenwich Observatory USAF/NOAA data showing that the magnetic equator was located a few degrees south of the heliographic equator.

\FloatBarrier

%

\FloatBarrier

\begin{figure} [!ht]
\centerline{\includegraphics[width=0.7\textwidth,clip=]{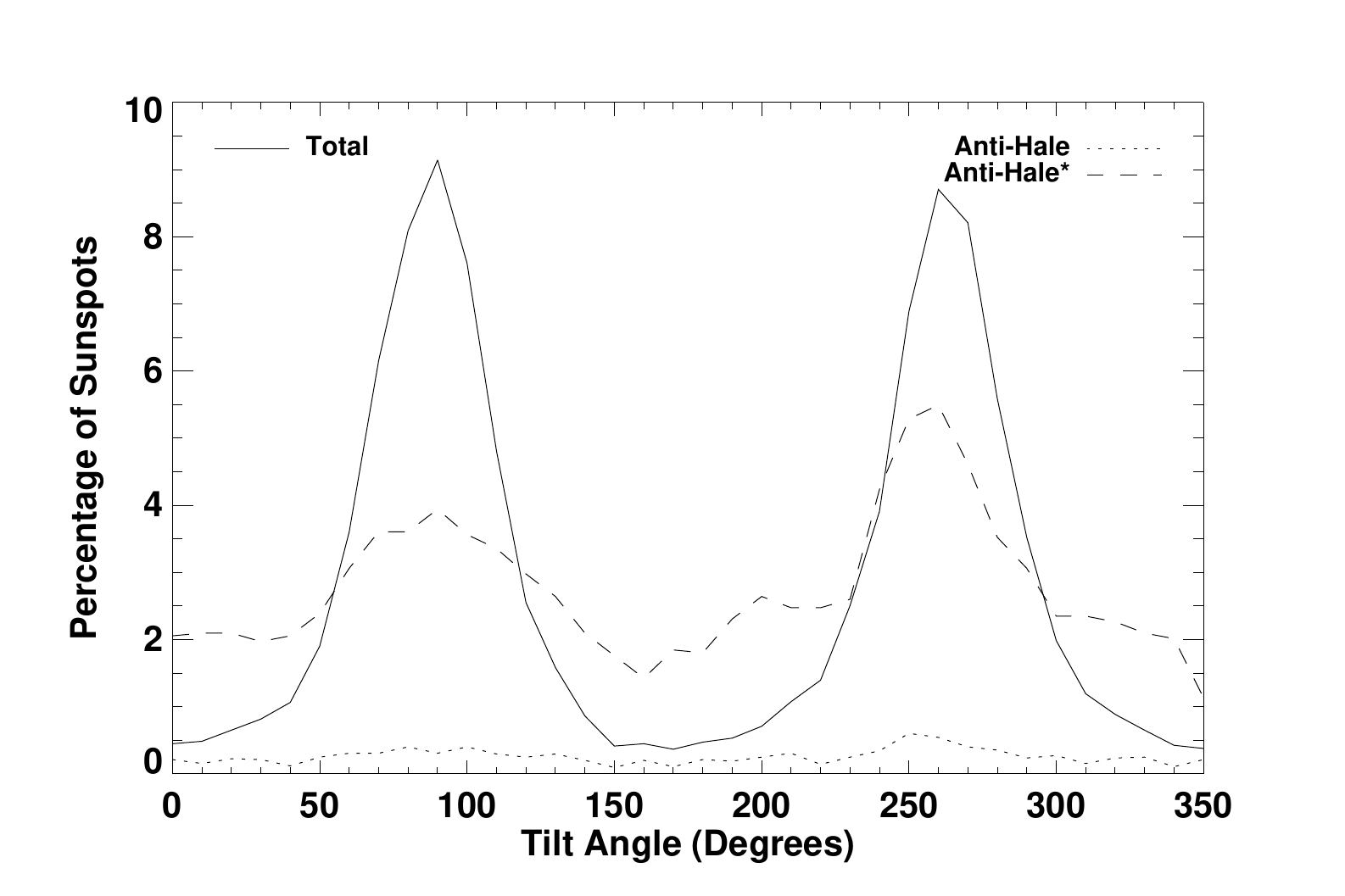}}
\caption{Tilt angles (0 - 360\degree) binned in 10$^{\circ}$ intervals. Distribution of all sunspot tilt angles (solid), anti-Hale numbers normalized by the total of all tilt angles (dot), and anti-Hale* normalized by the total of anti-Hale only (dash).}\label{tiltdistr}
 \end{figure}

Tilt angles with polarity (0 - 360\degree) were binned in 10\degree\ bins.  The number of sunspots for each bin is shown in Figure \ref{tiltdistr} as total (solid).  We plotted anti-Hale tilt angles normalized by the total number of sunspots (dot) and normalized by the total number of anti-Hale spots (dash). Anti-Hale tilt angles are part of a broader distribution of all tilt angles and show a weak dependence on being tilted 180\degr\ from their expected Joy's law angle.  The reason for this could be two-fold: (1) that east-west orientations of active regions are preferred and (2) the active regions contributing to the 90\degree\ and 270\degree\ peaks are the late-cycle, near-equator sunspot groups that are classified as anti-Hale because the magnetic equator is offset from the heliographic equator.

%

\begin{figure} [!ht]
\centering
\subfigure[]{\includegraphics[width=0.7\textwidth,clip=]{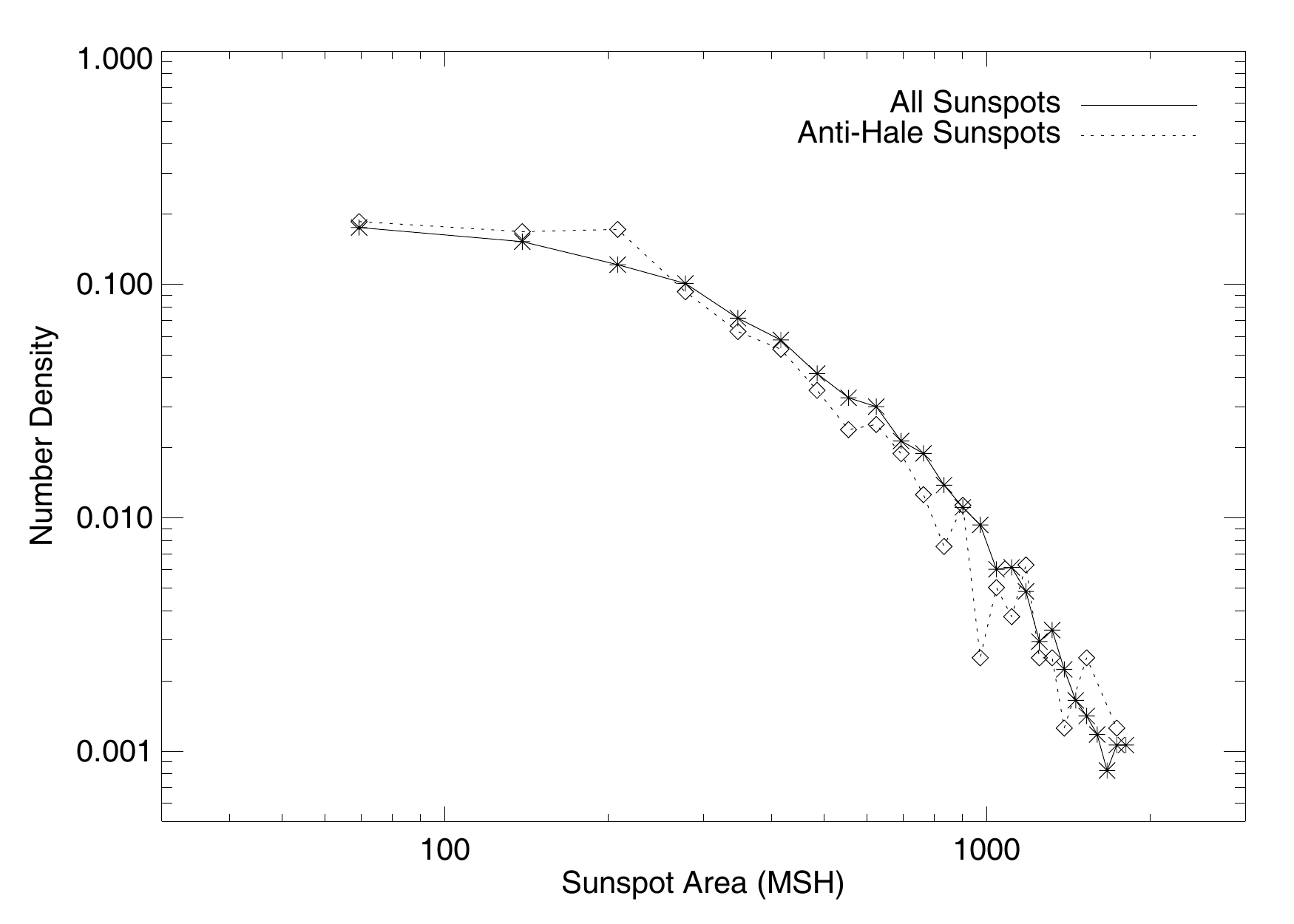}}
\subfigure[]{\includegraphics[width=0.7\textwidth,clip=]{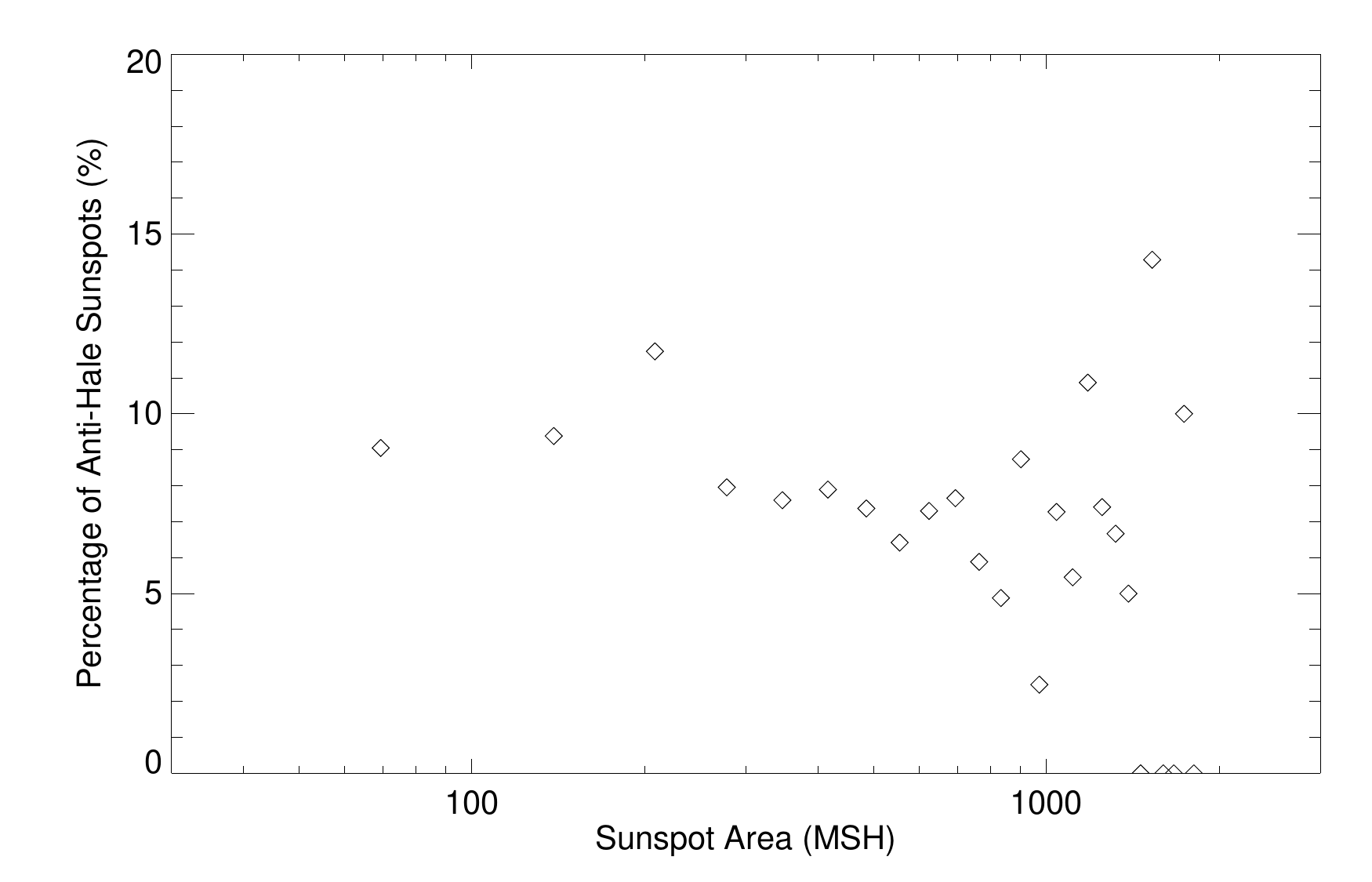}}
\caption{(a) Size distribution function of sunspot magnetic areas in micro-hemispheres (MSH) for all sunspot groups (asterisk, solid) and anti-Hale sunspots (diamond, dot) on a log-log scale. (b) Percentage of anti-Hale sunspots as a function of sunspot magnetic area in micro-hemispheres (MSH) plotted to log-normal scale.}\label{ahdist}
 \end{figure}

\citet{liu12} located sunspots on magnetograms to determine sunspot magnetic area in micro-hemispheres (MSH).  Figure \ref{ahdist}(a) shows the size distribution function of sunspot area for all bipolar sunspots (asterisk, solid) and anti-Hale sunspots (diamond, dotted).  The distributions for all sunspots and anti-Hale sunspots are similar to the log-log shape reported by \citet{bau05} and \citet{bog88}.  The percentage of anti-Hale spots for any given size is roughly 10\% as seen in Figure \ref{ahdist}(b).

%

\FloatBarrier

\section{Equatorial Regions}

\begin{figure}[!ht]
\centerline{\includegraphics[width=0.7\textwidth,clip=]{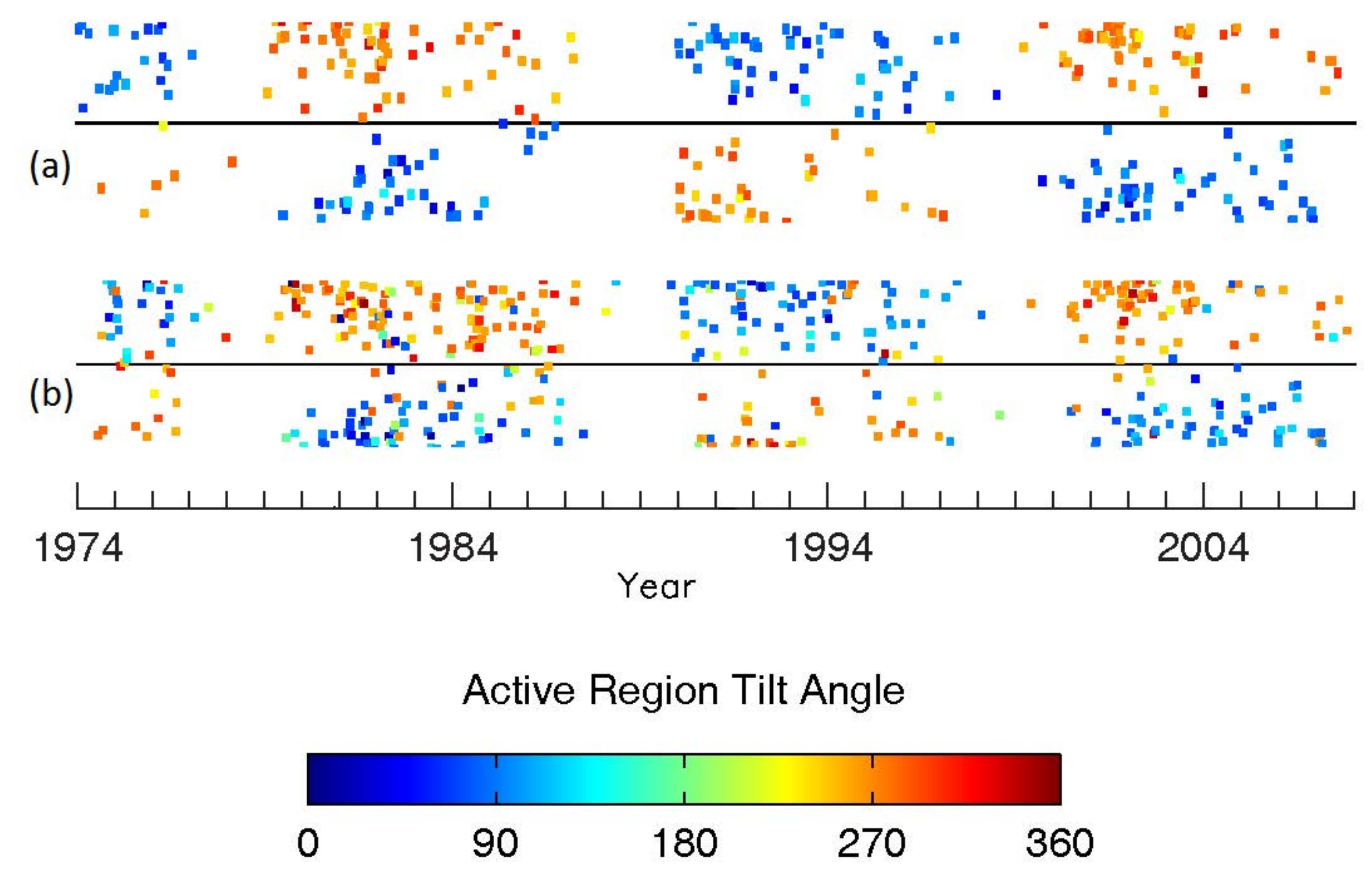}}
\caption{Bipolar sunspot regions within 5\degr\ of equator from 1974 to 2007. (a) DPD data are incomplete at the time of publication, most significantly from 1980 to 1985. (b) Li \& Ulrich data.  Panels (a) and (b) are cropped from Figures \ref{bflydpd} and \ref{bflyli}, respectively.}\label{bflyx2}
 \end{figure}

We examine the tilt angle of sunspots close to the equator, comparing DPD tilts without anti-Hale information (Figure \ref{bflyx2}(a)) to Li \& Ulrich tilt angles that include anti-Hale (Figure \ref{bflyx2}(b)). We focus attention within 5\degr\ of the equator between 1974-2007.  No bipolar regions were within 5\degr\ of the equator for the Li and Ulrich data after solar minimum in 2008 January. Although the DPD data are incomplete at the time of publication and some sunspots are present in Figure \ref{bflyx2}(b) that are not found in Figure \ref{bflyx2}(a), it can be noted that tilt angles near the equator in Figure \ref{bflyx2}(a) are at times miscalculated because the algorithm utilized in the DPD data set does not allow for any tilt angles outside the $\pm90$\degr\ range. In Figure \ref{bflyx2}(b), one can find anti-Hale sunspots in any given cycle shown as data points that are 180\degr\ opposite the dominant color. Within 5\degr\ of the equator, 65 out of 470 (13.8\%) bipolar sunspot regions are anti-Hale.  This percentage is slightly lower if sunspots at all latitudes are included.  From 1974 to 2012, 8.4\% of all Li \& Ulrich tilt angles are anti-Hale.  We therefore assume DPD data have incorrect tilt angles since anti-Hale are not recorded as such.

A southward-deflected magnetic equator in relation to the heliographic equator for the past 40 yr \citep{zol09}, also seen in Figure \ref{ahtot}(c), causes an increase in the number of sunspots that are categorized as anti-Hale late in cycle because northern hemispheric polarity sunspots are appearing south of the heliographic equator.  That the magnetic equator is shifted in relation to the heliographic equator can explain the increase in percentage of anti-Hale spots late in solar cycle.   Others \citep{mci2013, nor14} have investigated the asymmetry of the hemispheres by studying photospheric magnetism and also found that the northern hemisphere has been leading the southern hemisphere in Cycle 24.  As mentioned in the discussion of Figure \ref{tiltdistr}, equatorial regions that appear as anti-Hale, because they are northern hemispheric regions emerging slightly south of the heliographic equator contribute to the slight peaks at 90\degr\ and 270\degr\ in the anti-Hale tilt angle distributions.

In addition to assigned tilt angle range, other differences between DPD data and those of Li \& Ulrich become apparent in Figure \ref{bflyx2}.  Data-point density differences could be attributed to weather at each of the sites precluding agreement on any given day. The DPD conversion to digitized data, although nearly complete, is not finished.  DPD data identified the active regions by NOAA number while Li and Ulrich relied on Mount Wilson numbering until 1990, switching to NOAA in 1991, which could explain the higher number of data points before 1991 in  Figure \ref{bflyx2}(b) when compared to  Figure \ref{bflyx2}(a).

\section{Other Efforts to Include Tilt Angle in the Context of the Butterfly Diagram}

\begin{figure} [!ht]
\centering
\subfigure[]{\includegraphics[width=0.65\textwidth,clip=]{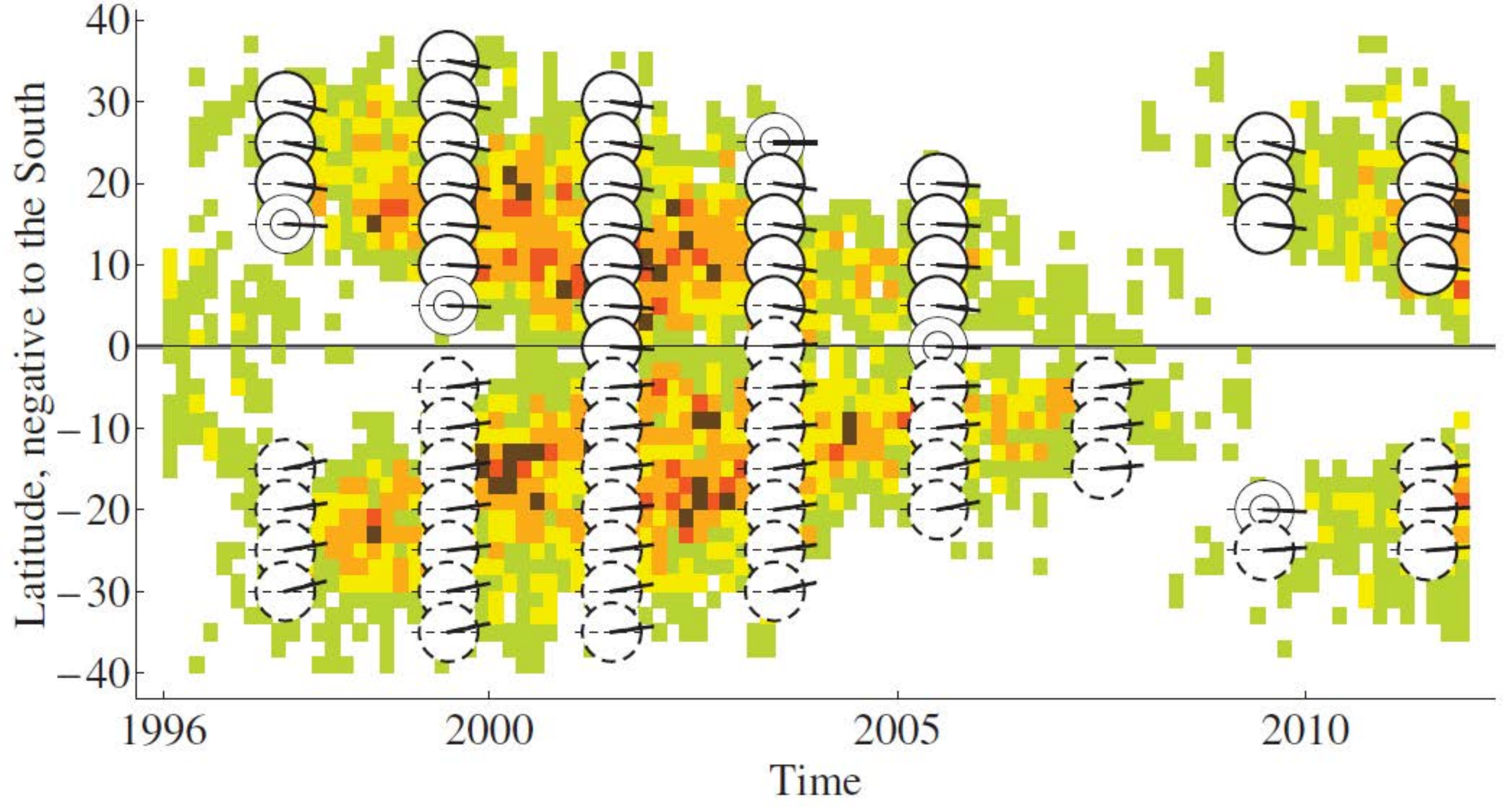}}
\subfigure[]{\includegraphics[width=0.65\textwidth,clip=]{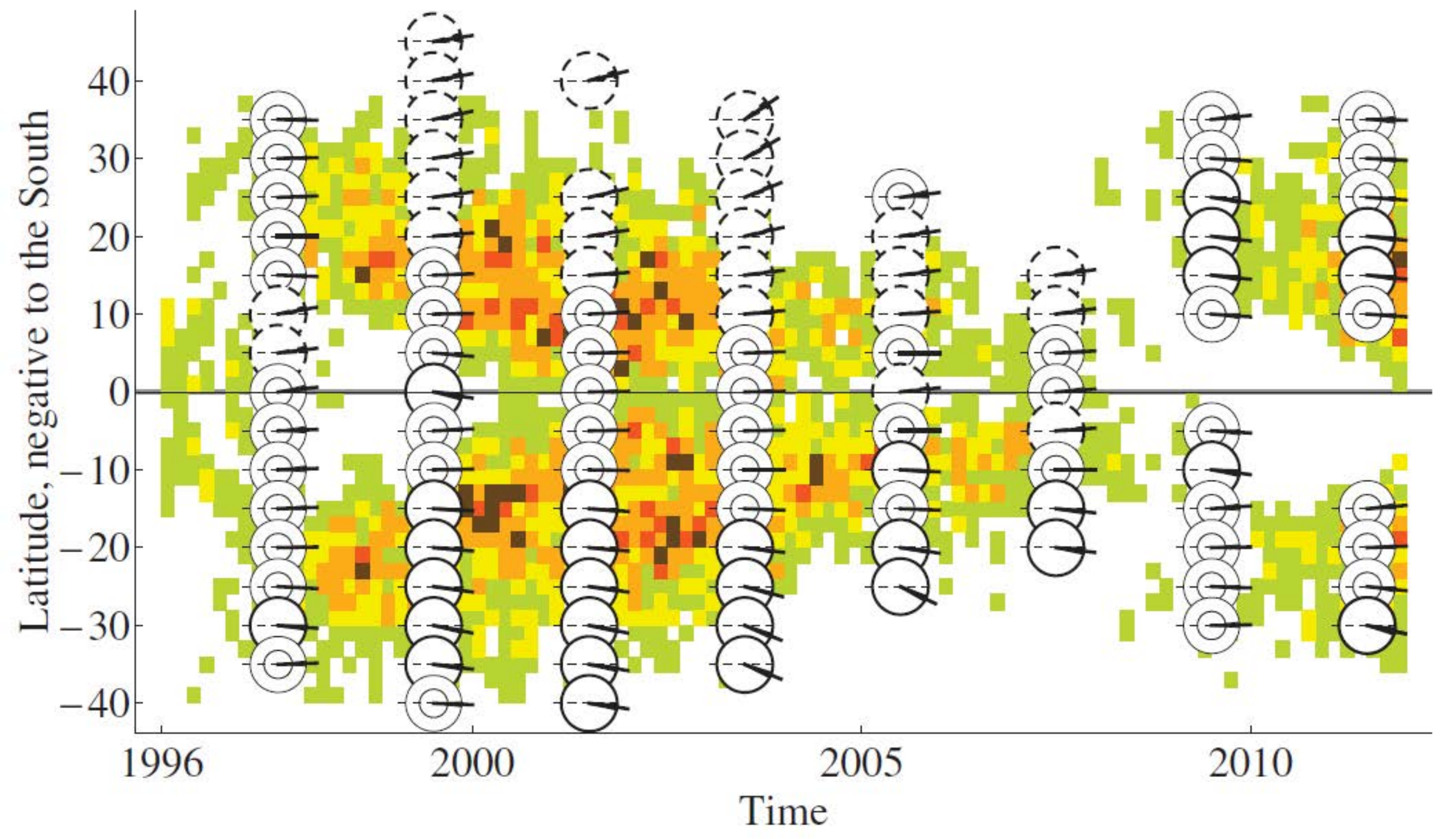}}
\caption{Figure of Solar Cycle 23 ($B_{min}$ = 10 G) reproduced with permission from \citet{tla13}. The underlying butterfly diagram with color representing area does not change in (a) and (b).  (a) Tilt angles are represented for larger sunspots with areas S $>$ 300 MSH from weighted MDI data. (b)  Tilt angles are represented for smaller sunspots with  areas 50 $<$ S $<$ 300 MSH for non-weighted MDI data.  Circles indicate mean tilt angle.  Solid (dashed) circles indicate positive (negative) mean tilt angle.  Zero tilt does not belong to the confidence interval, unless a double circle is present.  Tilt is positive (negative) if the tilt is clockwise (counterclockwise), regardless of hemisphere.}\label{tla13}
 \end{figure}

Plotting quantities as a function of latitude and time has illustrated many physical processes of the sunspot cycle.  Because tilt angle as a function of latitude is noisy as seen in Joy's law, it is unclear whether including tilt angle in the butterfly diagram can be useful.  Previous efforts include \citet{tla13}, who used weighted MDI data to plot tilt angle information over butterfly diagrams that used color to indicate sunspot area.  Large sunspot areas were defined as larger than 300 millionths of the solar hemisphere (MSH) and small areas as between 50 and 300 MSH.  In Figure \ref{tla13}(a), mean tilt angles for large sunspot areas are oriented as expected with positive (negative) mean tilt angles in the northern (southern) hemisphere. Data represented by double circles have a mean tilt value that is indistinguishable from zero.  Smaller sunspot areas (Figure \ref{tla13}(b)) have mean tilt angles at high latitudes mostly oriented away from what we would expect. Mean tilt values are nosier in the smaller sunspots (Figure \ref{tla13}(b)) and are not well determined at the beginning of cycles or near the equator.  Perhaps it is only useful insomuch that readers are able to understand that Joy's law is not well behaved or statistically easy to recover for a single hemisphere and solar cycle.  \citet{tla13} claims that these results are indicative of two distinct dynamo processes occurring, one that generates large sunspots and another that generates small sunspots. Overplotting tilt angle on the butterfly diagram was productive in their efforts after separating the large and small sunspots.

It appears that there is a contradiction in the literature, as follows.  \citet{kos08} do not find a dependence of tilt angle value on sunspot flux. (Note that sunspot flux and size are highly correlated.)  \citet{web13} simulations show that the tilt angle scatter increases for lower flux regions but the mean tilt angle does not vary significantly with flux. \citet{jia14} present Kodaikanal and Mt. Wilson Observatory tilt angle data binned according to sunspot size.  They find that ``the average tilt angles have a weak trend to increase with the sunspot group size, while the standard deviations significantly decrease" with sunspot group size.  However, \citet{tla13} find that smaller sunspots show more scatter and consistently have average tilt angles that are anti-Joy (not anti-Hale).  The contradiction may in part have its source in the different data used to determine group size: \citet{tla13} uses MDI magnetograms to determine area while \citet{jia14} use white-light intensity from ground-based observations.   \citet{kos08} also use MDI magnetograms, so the disagreement between \citet{tla13} and  Kosovichev and Stenflo is difficult to understand.

We disagree with  \citet{tla13} as we do not think there are two distinct dynamo processes occurring.  We also disagree with \citet{kos08} that anti-Hale spots indicate the presence of oppositely directed toroidal bands occurring simultaneously in the same hemisphere.  Rather, we agree with \citet{web13} simulations showing that the convective flows interact with the rising thin flux tubes to produce anti-Hale regions.  \citet{web13} find ``that 6.9\% emerge with polarities that violate Hale$'$s Law, in comparison to the $\approx4\%$ as found via observations" of \citet{wan89} and \citet{ste12}. We agree with \citet{web13} observations that the anti-Hale spots ``arise as a result of flux tubes emerging in the opposite hemisphere from which they originated, or as a result of the flux tube becoming so distorted by convection that the legs of the emerging loop can become reversed."

X. Sun and T. Hoeksema (in preparation) are using tilt angle information from the DPD data set to illustrate how flux transport on the solar surface reverses the polarity of the Sun's poles. 

%

\FloatBarrier

\section{Discussion and Conclusions}

We compared the statistics of a sunspot tilt angle catalog that utilizes magnetic polarity information to assign tilt angles from $0-360$\degr\ (Mount Wilson and MDI, Li \& Ulrich 2012) to a traditional tilt angle catalog in which the range of values is $\pm90$\degr\ (DPD).  Because historical tilt angle databases have not been capable of including anti-Hale information due to the lack of magnetograms or the unwillingness to incorporate the magnetogram information into a functional database, we paid particular attention to the anti-Hale statistics of bipolar sunspot regions.  We summarize our findings as follows.
 
\begin{enumerate}
\item   We find that 8.4\% $\pm$ 0.8\% of sunspot groups are anti-Hale from 1974-2012 as recorded in the Li \& Ulrich data, so we assume DPD data have incorrect tilt angles for these regions since anti-Hale values are not possible in the DPD value range.   The number of anti-Hale sunspots were found to be 9.0\%, 8.7\%, and 7.2\% of the total number of sunspot groups in Solar Cycles 21, 22, and 23, respectively.  Our reported 8.4\% value of anti-Hale regions is higher than previous studies.  This could simply be due to \citet{wan89} and \citet{ste12} using active regions that were not sunspots whereas we only use sunspots.   The number of anti-Hale regions at any given time is simply a fraction (see Figure \ref{ahtot}(b)) of the total number of sunspots present, excepting the end of each cycle when spots are very near the equator (see point 4, this section) and the expected polarity or source hemisphere is unknown.   

\item The average latitudes of anti-Hale regions are the same as all other sunspots for any given time in a solar cycle, meaning the average latitude of anti-Hale spots becomes more equatorward as the cycle progresses.  The size distribution of anti-Hale sunspots is the same as the log-normal distribution of all sunspots.  No area preferences emerge for anti-Hale regions.  

\item We find that anti-Hale are just part of a broader distribution of tilt angles, possibly as a result of convective zone turbulence as previously proposed by many researchers.  However, the misclassification of anti-Hale near the Equator is due to the heliographic equator not aligning with the magnetic equator.  This is visible in the tilt angle distribution as a slight tendency for the anti-Hale regions to be tilted 180\degree\ from their expected Joy's law tilt angle.

\item Joy's law cannot be observed by eye in the tilt-butterfly diagrams (Figures \ref{bflyli} and \ref{bflydpd}) but must be teased out statistically by averaging over significant periods of time.  This is evident by the lack of a smooth gradient in the color representing tilt angle in the butterfly wings when examining one hemisphere and one cycle. Even after averaging and binning, the trend in tilt angle shows discontinuous behavior in that the mean or median tilt angle will decrease on average for a period of years then increase again suddenly.   This may be related to distinct dynamo waves that occur within the solar cycle (Ternullo, 2007).

\item Sunspots very near the equator are often assigned incorrect tilt angles due to the magnetic equator being offset a few degrees.  For example, sunspots that have a northern hemispheric magnetic polarity and appear just below the heliographic equator (presumably because the northern hemisphere is ahead in the sunspot cycle and has reached the Equator first) are assigned tilt angles as if they are produced from the southern hemisphere polarity.  Figure \ref{ahtot}(a) shows an increase in anti-Hale near the end of each solar cycle.  Within 5\degr\ of the equator, 65 out of 470 (13.8\%) bipolar sunspot regions are anti-Hale.   The end of Solar Cycle 20 produced 14.8\% anti-Hale when sunspot activity is concentrated near the equator.  

Example: an active region (NOAA11987/HARP3784) straddles the equator but its central latitude is calculated to be 2\degr\ in the southern hemisphere on 2014 February 24 (Figure \ref{ar}(a)).  DPD reported a tilt angle on that date to be -15.26$^{\circ}$.  According to our definition, shown in Figure \ref{sc24ex}, the tilt angle should be 15.26$^{\circ}$.    The magnetogram in Figure \ref{ar}(b) shows a magnetic orientation consistent with a northern hemisphere BMR for this solar cycle. It is probably that this region, slightly south of the equator, originated from magnetic dynamo activity in the northern hemisphere since the Northern hemisphere is leading the southern hemisphere and reached the equator first.  Conversely, this region could have originated from dynamo action in the southern hemisphere with polarity orientation and tilt angle opposite from that anticipated via Joy's law and Hale's law.  We suggest the former is more plausible.  We propose assigning this region to the northern hemisphere based on tilt and polarity, with a caveat regarding latitude.  Also note that this active region has a significant tilt angle, not a zero tilt angle as predicted by many versions of Joy's law that forces the tilt to be zero at the equator.  The practice of forcing Joy's law to zero at the equator is not supported by observations. Doing so makes a huge difference in the slopes as reported in the literature and we find that there are, as often as not, BMRs with significant tilts at the equator.

%
\begin{figure}  [!ht]
\centerline{\includegraphics[width=0.6\textwidth,clip=]{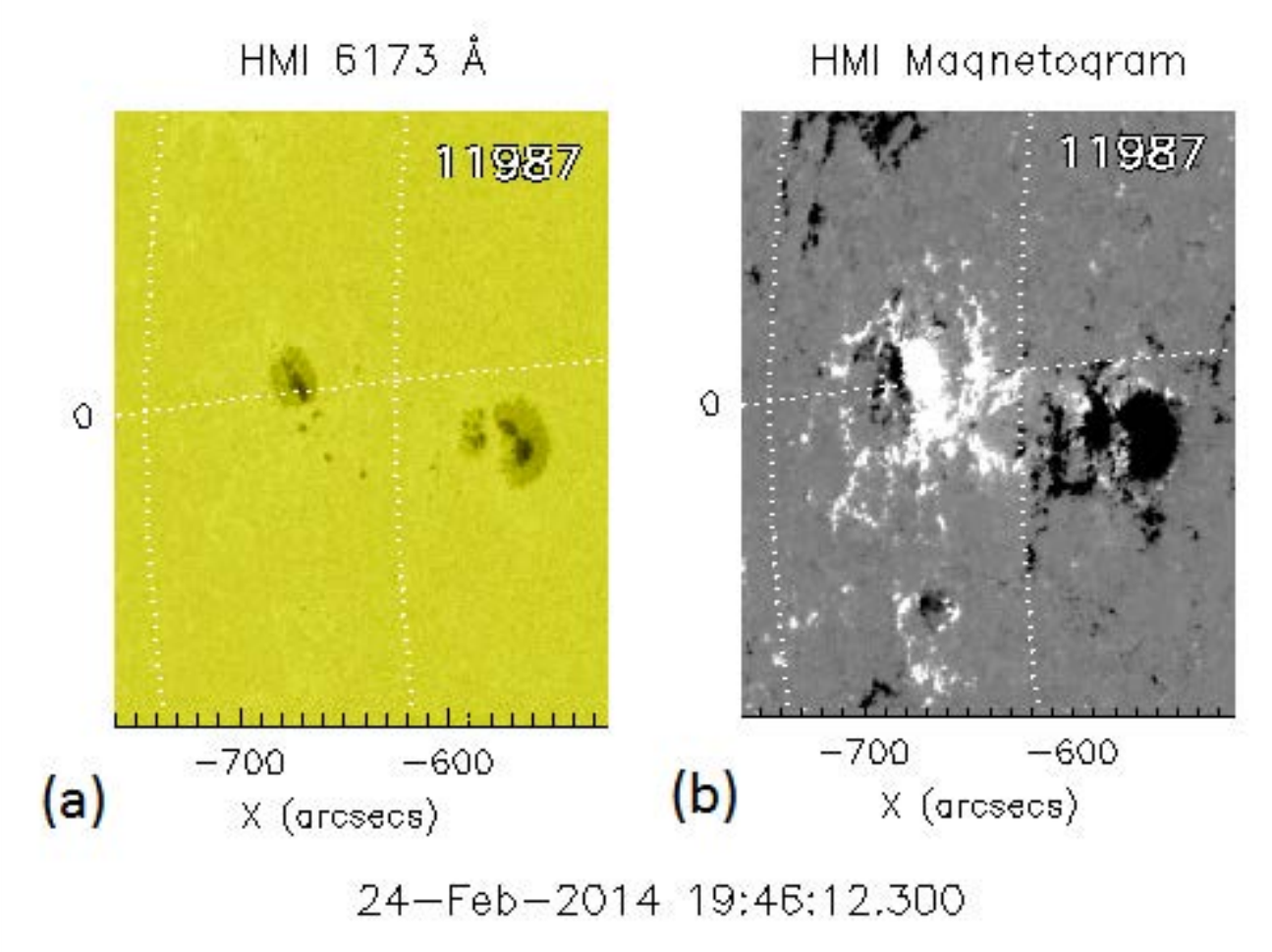}}  
\caption{(a) White light intensity image and (b) magnetogram on 2014 February 24 of bipolar sunspot group NOAA11987 straddling the equator.  The leading polarity of NOAA11987 is indicative of a northern hemisphere orientation for this solar cycle.  DPD reported a tilt of umbral activity as -15.26$^{\circ}$ whereas a northern hemisphere bipolar region of this orientation would be reported with a positive tilt angle.} \label{ar}
 \end{figure}

\item The tilt angles of sunspot groups and associated scatter in the tilt angle values are crucial for the build up and reversals of the polar fields in surface flux transport simulations \citep{cam10, cam12}.  Recent efforts by \citet{jia14} have shown that the tilt angle scatter constitutes a significant random factor in the variability of cycle strength.  Therefore, we anticipate that the inclusion of the 8.4\% of anti-Hale sunspots in surface flux transport models may allow for even greater cycle to cycle amplitude variability.

\end{enumerate}

\FloatBarrier

From points 1 and 2, we conclude that the physical processes that produce anti-Hale regions are the same processes that produce sunspots obeying Hale's polarity rules.  Knowing that tilt angles have a high scatter, we conclude that the tails of the tilt angle distribution function are quite wide and therefore 8.4\% of all spots can have angle $\pm90$\degr\ from the expected orientation.   Of course, many questions about tilt angles and anti-Hale activity remain unanswered.  There is, as always, more work to do.       



\FloatBarrier

\end{document}